\documentclass[twocolumn,twocolappendix]{aastex63}
\usepackage[utf8]{inputenc}
\usepackage{newtxtext,newtxmath}
\usepackage[T1]{fontenc}
\usepackage{ae,aecompl}
\usepackage{comment}

\usepackage{graphicx,tabularx}	
\usepackage{amsmath}	
\usepackage{amssymb}	
\usepackage{booktabs}
\usepackage[inline,shortlabels]{enumitem}
\setlist{itemjoin* = { and\enspace}}
\usepackage{color}
\usepackage{multirow}
\usepackage[figuresright]{rotating}
\newcommand{\hi}{H{~\sc i}}
\newcommand{\hii}{H{~\sc ii}}

\received{XXX}
\revised{XXX}
\accepted{XXX}

\submitjournal{ApJ}

\shorttitle{Reconstruction of Initial Density Field}
\shortauthors{Zhou \& Mao}

\graphicspath{{./}{figures/}}

\begin{document}
\title{Reconstruction of Cosmological Initial Density Field with Observations from the Epoch of Reionization}

\correspondingauthor{Meng Zhou, Yi Mao}
\email{zhoum18@mails.tsinghua.edu.cn (MZ), ymao@tsinghua.edu.cn (YM)}

\author[0000-0002-2744-0618]{Meng Zhou} 
\affiliation{Department of Astronomy, Tsinghua University, Beijing 100084, China}
\affiliation{Department of Computer Science, University of Nevada, Las Vegas, Las Vegas, NV 89154, USA}

\author[0000-0002-1301-3893]{Yi Mao} 
\affiliation{Department of Astronomy, Tsinghua University, Beijing 100084, China}

\begin{abstract}

Initial density distribution provides a basis for understanding the complete evolution of cosmological density fluctuations. While reconstruction in our local Universe exploits the observations of galaxy surveys with large volumes, observations of high-redshift galaxies are performed with a small field of view and therefore can hardly be used for reconstruction. Here we propose to reconstruct the initial density field using the \hi\ 21~cm and CO line intensity maps from the epoch of reionization. Observations of these two intensity maps provide complementary information of the density field --- the \hi\ 21~cm field is a proxy of matter distributions in the neutral regions, while the CO line intensity maps are sensitive to the high-density, star-forming regions that host the sources for reionization. Technically, we employ the conjugate gradient method and develop the machinery for minimizing the cost function for the intensity mapping observations. Analytical expressions for the gradient of cost function are derived explicitly. We show that the resimulated intensity maps match the input maps of mock observations using semi-numerical simulations of reionization with an rms error $\lesssim 7\%$ at all stages of reionization. This reconstruction is also robust at the same level of accuracy when a noise at the level of $\lesssim 1\%$ of the standard deviation is applied to each map. Our proof-of-concept work demonstrates the robustness of the reconstruction method, thereby providing an effective technique for reconstructing the cosmological initial density distribution from high-redshift observations. 

\end{abstract}

\keywords{Reionization (1383), H I line emission (690), CO line emission (262), Initial conditions of the universe (795), Line intensities (2084)}

\section{Introduction}

In the standard theory of cosmology (see, e.g.\ \citealt{2010gfe..book.....M}), initial density perturbations of our Universe were generated by quantum fluctuations and then amplified during the inflationary era. A hierarchy of inhomogeneous structures was formed from small scales (e.g.\ galaxies and halos) to large scales (e.g.\ filaments and voids) thanks to gravitational instability. Bridging theories and observations, statistical observables are usually employed to quantify the measurements, which nevertheless are subject to cosmic variance unavoidably. Given observational data of large-volume galaxy surveys in our local Universe, reconstruction of cosmological initial density field is a solution to avoid the cosmic variance, in that galaxies and halos from simulations can be compared directly to their counterparts in observations if the initial conditions are all known. Given the input observations of galaxy surveys, \citet{2013MNRAS.432..894J,2013ApJ...772...63W} sampled the posterior distribution function of the initial density field with the Hamiltonian Monte Carlo (HMC) method and then reconstructed the initial density field accurately in real space  \citep{2013MNRAS.432..894J} or in Fourier space \citep{2013ApJ...772...63W}.

Looking deeper into the Universe, observations of high-redshift galaxies by, e.g.\ the James Webb Space Telescope, are performed with a small field of view \citep{2023MNRAS.526.2542C} and therefore can hardly be used for reconstruction. However, line intensity mappings \citep{2022A&ARv..30....5B} emerge as a promising cosmological probe with a large field of view, generically. In particular, during the epoch of reionization (EoR), the tomographic mapping of the 21~cm line intensity, due to the hyperfine transition of atomic hydrogen, promises to make a 3D image that reveals the evolutionary history of the Universe from being neutral to ionized because \hi\ 21~cm line is a proxy of matter density in the neutral regions. Nevertheless, almost no 21~cm signal can be emitted from the ionized regions (aka \hii\ bubbles). Thus the matter densities in the ionized regions can only be inferred indirectly from their impact on the morphology of \hii\ bubbles in the 21~cm maps. In the inside-out scenario of reionization, the ionized regions started to form surrounding the first galaxies at high-density regions on average. To fill the gap of ionized regions in the 21~cm maps, therefore, the intensity mappings of molecular lines, e.g.\ CO(1-0) \citep{2011ApJ...728L..46G,2011ApJ...741...70L} and C{~\sc ii} line \citep{2012ApJ...745...49G,2015ApJ...806..209S,2019MNRAS.485.3486D} that are good tracers of high-density, star-forming regions that host the sources of reionization, serve as complementary cosmological probes during the EoR, e.g.\ they are anti-correlated with the 21~cm signal statistically \citep{2015aska.confE...4C}. 

In this paper, we propose for the first time to reconstruct the cosmological initial density field using future intensity mappings of the \hi\ 21~cm line and CO(1-0) line from the EoR. Technically, we will build new machinery for minimizing the cost function for the intensity mapping observations, and apply it to the EoR observations with sophisticated derivations of the analytical expressions for the gradient of cost function.

The rest of this paper is organized as follows. We introduce our method in Section~\ref{methodology}, show the main results in Section~\ref{result}, and make concluding remarks in Section~\ref{conclusion}. Some technical details are left to Appendix~\ref{strategy} (on the optimization of reconstruction) and Appendix~\ref{app:mean} (on the effect of subtracting the mean from signals).
\section{Methodology}
\label{methodology}

The initial density reconstruction involves two ingredients: (1) a framework that finds the most likely configuration of the initial density field from which the resimulated maps match the input maps of observation, and (2) an underlying theoretical model or simulation tool that connects the initial density field with the simulated maps. For instance, the reconstruction from the galaxy survey observation employs the HMC method as the matching algorithm and the second-order Lagrangian perturbation theory (in \citealt{2013MNRAS.432..894J}) or modified Zel'dovich approximation (in \citealt{2013ApJ...772...63W}) for evolving the initial density field to the matter distribution at $z=0$. In this paper, for the reconstruction at the time of EoR, we employ the {\it conjugate gradient} method\footnote{We also made an attempt to apply the HMC method to the reconstruction from the EoR. However, we can not optimize the HMC to a satisfactory level of accuracy (with the reconstruction error $< 10\%$) within an affordable time of computation.} to minimize the cost function directly. Also, as a first attempt for such a reconstruction, we employ the excursion set model of reionization (ESMR; \citealt{2004ApJ...613....1F}) as the underlying reionization theory to evolve the initial density field to the ionization field and the Zel'dovich approximation for the evolution of density fluctuations. Throughout this paper, we adopt a $\Lambda$CDM cosmology model with $\Omega_{\rm m} = 0.31$, $\Omega_{\rm b} = 0.048$, $h = 0.68$, and $\sigma_8 = 0.81$.

In \S\ref{subssec:CG}, a general framework for reconstruction will be constructed, assuming that an underlying theory for reionization and density fluctuations is available, and a specific implementation will be made in \S\ref{analytical gradients} (using the Zel'dovich approximation), \S\ref{subsec:ESMR} (modeling of the 21~cm and CO line intensity maps) and \S\ref{subsec:analytical-2} (using the ESMR), with a description of our simulation setup in \S\ref{subsec:sim}. 


\subsection{Reconstruction Framework} 
\label{subssec:CG}


Consider an initial density field $\delta^{\rm{ini}}_n$, where the subscript $n$ is the index of comoving cells in 3D real space. In this section, we also represent the initial density field as an $N$-dimensional vector $\boldsymbol{\delta}^{\rm ini}$, where $N$ is the total number of cells. 

Assume that the initial density field is evolved to the simulated maps, $T_{j,\,n}^{\,\rm{mod,\,coev}}$ in a coeval simulation box at a later observation time. Here, the superscript ``mod'' refers to the map from simulations or models, ``coev'' refers to the coeval box, and the subscript $j$ refers to the type of observation maps, e.g.\  in \S\ref{subsec:ESMR}, $j =$ ``21cm'' and ``CO'', which denote the 21~cm and CO(1-0) intensity maps, respectively. 

Note that the simulated maps usually have higher resolutions, in order to resolve nonlinear structures, than the maps of observation that are subject to instrumental effects. Thus we use $T_{j,\,\alpha}^{\,\rm{mod}}$ to denote the simulated maps that are smoothed to match the resolution in observation, where $\alpha$ is the index of comoving {\it coarse-grained} cells. 
\begin{eqnarray}
    \label{average}
    T^{\rm{mod}}_{j,\,\alpha} = C_{\rm trans}\sum_{n\in \alpha} T^{\rm{mod,\,coev}}_{j,\,n}\,. 
\end{eqnarray}
Here, ``$n\in \alpha$'' refers to the summation over the simulation cells (with the dummy index $n$) that are inside the coarse-grained cell with the index $\alpha$, and $C_{\rm trans}$ is a normalization factor that parametrizes the contribution of each simulation cell to the coarse-grained cell. We adopt a simple treatment of smoothing in Equation~(\ref{average}) for the purpose of the proof of concept. In principle, the smoothed simulated maps should be the convolution of the simulated maps with some window function or filters. More instrumental and/or observational effects can also be included in the forward simulation, which will be the focus of a follow-up paper. 

Our goal is to match the smoothed simulated maps $T_{j,\,\alpha}^{\,\rm{mod}}$ to the input maps of observation $T_{j,\,\alpha}^{\,\rm{inp}}$ as accurately as possible. For this purpose, we define the {\it cost function}, which is the negative logarithm of the likelihood, as a {\it functional} of the initial density field \citep{2013ApJ...772...63W}, 
\begin{eqnarray}
    \label{cost}
    \phi( \boldsymbol{\delta}^{\rm ini}) &=& C_{\rm {cost}}\sum_{j,\,\alpha}\frac{(T^{\,\rm{mod}}_{j,\,\alpha}-T^{\,\rm inp}_{j,\,\alpha})^2}{2(\sigma^{\rm N}_{j})^2}w_{j}\,, 
\end{eqnarray}
where $C_{\rm cost}$ is an overall normalization factor and $w_{j}$ is the weight between different types of observations that can be adjusted for optimization, $\sum_j w_j = 1$. Here we assume a white noise $\sigma^{\rm N}_{j}$ that sums up all noises in observations. In the ideal case with only cosmic signals and no noise, we take $\sigma^{\rm N}_{j}$ to be $0.1\%$ of the standard deviation in each mock map $j$. In principle, the framework in this paper can be readily extended to include more realistic effects of noises and systematics, which we will investigate in a follow-up paper. 

Reconstruction of the initial density is the process of minimizing the cost function (or equivalently maximizing the likelihood). If we regard the initial density field at each point $n$ as a free parameter, then this is a typical problem of multidimensional optimization. Multidimensional optimization is usually realized by performing one-dimensional optimization (i.e.\ line minimization) iteratively along different line directions. Here we introduce the conjugate gradient method \citep{10.5555/1403886} that is widely applied in multidimensional optimization. The conjugate gradient method is particularly computationally efficient by constructing a set of directions that are mutually conjugate, which speeds up the convergence. 

We start from a guess of initial density field $\boldsymbol{\delta}^{\rm ini}_0$ that is randomly picked from a multi-dimensional Gaussian distribution with the covariance matrix given by a modified linear power spectrum\footnote{The modified power spectrum is set to be the theoretical linear matter power spectrum multiplied by a factor of $0.01$. This overall reduction in the amplitude of density fluctuations, while maintaining the Gaussianity, can mitigate the likelihood of generating ill-conditioned points.}.  Suppose that the underlying theory of reionization and density fluctuations is differentiable, i.e.\ an analytical expression for the gradient $\boldsymbol{g} = \partial \phi/\partial \boldsymbol{\delta}^{\rm ini}$ can be derived from the theory, which is the focus of \S\ref{analytical gradients} and \S\ref{subsec:analytical-2}. For the initial guess, we evaluate the initial gradient $\boldsymbol{g}_0 = \partial \phi/\partial \boldsymbol{\delta}^{\rm ini}_0$ and set $\boldsymbol{h}_0 = -\boldsymbol{g}_0$ as the initial direction of line minimization. We then perform the iterations of the conjugate gradient until convergence is achieved. Assuming that $\boldsymbol{\delta}^{\rm ini}_i$, $\boldsymbol{g}_i$ and $\boldsymbol{h}_i$ are given from the previous iteration $i$, the algorithm in the $(i+1)$-th iteration is as follows. 

(1) Perform the line minimization along the direction $\boldsymbol{h}_i$ and find the minimum point $\lambda_{{\rm min},i}$ (see the detail below).

(2) Update the field $\boldsymbol{\delta}^{\rm ini}_{i+1} = \boldsymbol{\delta}^{\rm ini}_i+\lambda_{{\rm min},i} \boldsymbol{h}_i$.

(3) Update the gradient $\boldsymbol{g}_{i+1} = \partial \phi/\partial \boldsymbol{\delta}^{\rm ini}_{i+1}$. 

(4) Update the direction $\boldsymbol{h}_{i+1} = -\boldsymbol{g}_{i+1} +\gamma_i \boldsymbol{h}_i$, where $\gamma_i$ is the Polak-Ribiere variant \citep{polak1971computational}, $\gamma_i = (\boldsymbol{g}_{i+1}-\boldsymbol{g}_i)^{\rm T} \boldsymbol{g}_{i+1}/(\boldsymbol{g}_i^{\rm T} \boldsymbol{g}_i)$. 

(5) Go to the next iteration. 

To perform the line minimization, we follow the Dbrent method \citep{10.5555/1403886}, a modified version of Brent's method \citep{Brent1973}. Given the field $\boldsymbol{\delta}^{\rm ini}_i$ and the direction $\boldsymbol{h}_i$, the cost function $\phi(\boldsymbol{\delta}^{\rm ini}_i+\lambda \boldsymbol{h}_i)$ is a scalar function of $\lambda$, where $\lambda$ is a free parameter. Its derivative is  
\begin{equation}
    \frac{\partial \phi}{\partial \lambda} = \left.\frac{\partial \phi(\boldsymbol{\delta}^{{\rm ini}\,'})}{\partial \boldsymbol{\delta}^{{\rm ini}\,'}}\right|_{\boldsymbol{\delta}^{{\rm ini}\,'} = \boldsymbol{\delta}^{\rm ini}+\lambda \boldsymbol{h}_i} \cdot \boldsymbol{h}_i.
    \label{eq:lm}
\end{equation}
The Dbrent method finds the value of $\lambda$ at the minimum point, $\lambda_{{\rm min},i}$, by tightening a bracket of $\lambda$ values using the sign of the derivative evaluated with Equation~(\ref{eq:lm}) at the central point of the bracket.

\subsection{Analytical Expression for the Gradient with respect to the Initial Density}
\label{analytical gradients}

This subsection focuses on deriving an analytical expression for the gradient of the cost function with respect to the initial density field, $\partial \phi / \partial \delta^{\rm{ini}}_n$. 
The cost function is generally a function of the evolved density field $\boldsymbol{\delta}^{\rm evol}$ that is in turn a function of the initial density field $\boldsymbol{\delta}^{\rm ini}$. Both $\boldsymbol{\delta}^{\rm evol}$ and $\boldsymbol{\delta}^{\rm ini}$ are $N$-dimensional vectors. The superscripts ``evol'' and ``ini'' imply the evolved and initial density field, respectively. 
Thus the gradient of the cost function is written using the chain rule as   
\begin{equation}
\frac{\partial \phi}{\partial \delta^{\rm{ini}}_n} = \sum_m\frac{\partial \phi}{\partial \delta_m^{\rm evol}}\frac{\partial \delta_m^{\rm evol}}{\partial \delta_n^{\rm ini}}\,. 
\label{likelihood derivative1}
\end{equation}

In this paper, we employ the Zel'dovich approximation for the evolution of density fluctuations, i.e.\ for connecting the initial density field $\boldsymbol{\delta}^{\rm ini}$ with the evolved density field $\boldsymbol{\delta}^{\rm evol}$. Zel'dovich approximation assumes that the separation between the comoving coordinates of a particle and its Lagrangian coordinates is linear to the time-independent displacement. Given the initial position $\boldsymbol{x}_p^{\rm{ini}}$ of the $p$-th particle, its final position $\boldsymbol{x}_p^{\rm{evol}}$ is given by  
\begin{equation}
    \boldsymbol{x}_p^{\rm{evol}} = \boldsymbol{x}_p^{\rm{ini}}+(D^{\rm evol}-D^{\rm ini})\,\boldsymbol{v}_p \,,
    \label{ZA}
\end{equation}
where $D^{\rm evol}$ and $D^{\rm ini}$ are growth factors at the redshifts of the evolved and initial density fields, respectively. The displacement $\boldsymbol{v}_p$ can be written as 
\begin{equation}
    \boldsymbol{v}_p = \frac{1}{N}\sum_q\frac{\boldsymbol{k}_q \sqrt{-1}}{k_q^2}e^{2\pi pq\frac{\sqrt{-1}}{N}}\sum_l\delta^{\rm ini}_le^{-2\pi lq\frac{\sqrt{-1}}{N}}\,.
    \label{Displacement}
\end{equation}
where $\boldsymbol{k}_q$ is the $q$-th wave vector.\footnote{The convention of the Fourier Transform (from $\boldsymbol{x}$-space to $\boldsymbol{k}$-space) and the inverse Fourier Transform (from $\boldsymbol{k}$-space to $\boldsymbol{x}$-space) is as follows. 
$\mathcal{F}(f) = g_k = \hat{C}\sum_{j=0}^{N-1}f_j \exp{(-2\pi jk \sqrt{-1}/N)}$, and $ \mathcal{F}^{-1}(g) =  f_j = C\sum_{k=0}^{N-1}g_k \exp{(2\pi jk \sqrt{-1}/N)}$, where $j$ and $k$ are labels in the position space and in the Fourier space, respectively. We write the three-dimensional transform in a one-dimensional format for simplicity in this paper.
Here $C$ and $\hat{C}$ are normalization factors that satisfy the relation $C\hat{C} = 1/N$. 
}

We smooth the particles to mesh with the Cloud-in-Cell (CIC) algorithm. We assume a cubic particle as the mass of $(1+\delta^{\rm{ini}})$ with the same volume of a mesh cell. Therefore, the evolved density is 
\begin{equation}
    1+\delta_m^{\rm{evol}} = \sum_p(1+\delta^{\rm{ini}}_p) \, W_{\rm{CIC}}(\boldsymbol{x}_p^{\rm{evol}}-\boldsymbol{x}_m)\,,
\end{equation}
where $W_{\rm{CIC}}(\boldsymbol{x})$ is the CIC window function, $\boldsymbol{x}_m$ is the position of the $m$-th mesh cell for the evolved density field at the final redshift, $\boldsymbol{x}_p^{\rm{evol}}$ is the final position of the $p$-th particle at the final redshift. The $p$-th particle is the $p$-th mesh cell for the initial density field at the initial redshift.

Thus the partial derivative $\partial \delta^{\rm{evol}}_m / \partial \delta^{\rm {ini}}_n$ can be written as 
\begin{equation}
    \frac{\partial \delta^{\rm{evol}}_m}{\partial \delta^{\rm {ini}}_n} = W_{\rm{CIC}}(\boldsymbol{x}_n^{\rm{evol}}-\boldsymbol{x}_m)
+\sum_p(1+\delta^{\rm{ini}}_p) W^\prime_{\rm{CIC}}(\boldsymbol{x}_p^{\rm{evol}}-\boldsymbol{x}_m) \cdot \frac{\partial \boldsymbol{x}_p^{\rm{evol}}}{\partial \delta^{\rm {ini}}_n}\,,
\end{equation}
where 
\begin{equation}
    \frac{\partial \boldsymbol{x}_p^{\rm{evol}}}{\partial \delta^{\rm {ini}}_n} = \frac{1}{N}(D^{\rm evol}-D^{\rm ini})\sum_q\frac{\boldsymbol{k}_q \sqrt{-1}}{k_q^2}e^{2\pi (p-n)q\frac{\sqrt{-1}}{N}}\,.
\end{equation} 
Note that $W^\prime_{\rm{CIC}}(\boldsymbol{x})$ --- the derivative of the CIC window function with respect to $\boldsymbol{x}$ --- is a vector. 

Suppose that the underlying theory of reionization is differentiable, i.e.\ an analytical expression for the gradient $\partial \phi/\partial \boldsymbol{\delta}^{\rm evol}$ can be derived from the theory, which is the focus of \S\ref{subsec:analytical-2}. For simplicity, we define $A_m=\partial \phi/\partial \delta_m^{\rm evol}$ and $\boldsymbol{B}_p=\sum_m A_m(1+\delta^{\rm ini}_p)\,W^\prime_{\rm{CIC}}(\boldsymbol{x}_p^{\rm{evol}}-\boldsymbol{x}_m)$. 
Putting all together, the gradient of the cost function can be written as 
\begin{eqnarray}
\label{dphidd}
    \frac{\partial \phi}{\partial \delta^{\rm{ini}}_n} &=& \sum_m A_m W_{\rm{CIC}}(\boldsymbol{x}_n^{\rm{evol}}-\boldsymbol{x}_m)\nonumber\\
    & & \!\!\!\!\!\!\!\!\!\!\!\!\!\!\!\!\!\!\!\!\!\!+\frac{1}{N}(D^{\rm evol}-D^{\rm ini})\sum_{p,\,q} \boldsymbol{B}_p \cdot \boldsymbol{k}_q \,\frac{ \sqrt{-1}}{k_q^2}e^{2\pi (p-n)q\frac{\sqrt{-1}}{N}}\,.
\end{eqnarray}
Equation~(\ref{dphidd}) can be computed efficiently with the FFT. Basically, $\boldsymbol{B}_p$ and the first term in the RHS of Equation~(\ref{dphidd}) are just convolutions of $A_m$ and some kernels, so they can be calculated in Fourier space using the convolution theorem. Also, the second term in the RHS of Equation~(\ref{dphidd}) can be rewritten as follows, so it can calculated easily with the FFT. 
\begin{eqnarray}
    & & \frac{1}{N}(D^{\rm evol}-D^{\rm ini})\sum_q \frac{\boldsymbol{k}_q \sqrt{-1}}{k_q^2}e^{-2\pi nq\frac{\sqrt{-1}}{N}} \cdot \sum_p \boldsymbol{B}_p\, e^{2\pi pq\frac{\sqrt{-1}}{N}}\nonumber\\
    &=& (D^{\rm evol}-D^{\rm ini})\,\mathcal{F}^{-1}\left[\frac{\boldsymbol{k}_q \sqrt{-1}}{k_q^2}\cdot \mathcal{F}(\boldsymbol{B})_q^* \right]_{N-n}\,.\nonumber
\end{eqnarray}

\subsection{The 21~cm and CO Intensity Mappings}
\label{subsec:ESMR}



The EoR 21~cm brightness temperature $T^{\rm{mod,\,coev}}_{{\rm 21cm},\,n}$ at the comoving cell $n$ is given by the local neutral fraction $x_{{\rm HI},\,n}$ and the local overdensity $\delta_n^{\rm evol}$, 
\begin{equation}
\label{T21}
    T^{\rm{mod,\,coev}}_{{\rm 21cm},\,n} = c_{{\rm 21cm}}\,x_{{\rm HI},\,n}\,(1+\delta_n^{\rm evol})\,,
\end{equation}
where $c_{\rm 21cm} = 27\, \sqrt{[(1+z)/10]\,(0.15/\Omega_{\rm m} h^2)}\,(\Omega_{\rm b} h^2/0.023)$ in units of millikelvins. Here, all quantities are implicitly evaluated at a time during the EoR. In this paper, we focus on the limit where the spin temperature is much higher than the cosmic microwave background temperature, which is valid soon after reionization begins. As such, the dependence on spin temperature is neglected as in Equation~(\ref{T21}). 

We employ the ESMR to identify the ionized regions. Specifically, cells inside a spherical region are identified as ionized, if the number of ionizing photons in that region is larger than that of neutral hydrogen atoms,  or $f_{\rm coll}(\delta_{R,\,n}^{\rm evol},\, M_{\rm vir}) \ge \zeta^{-1}$. Here, $\zeta$ is the ionizing efficiency, $M_{{\rm vir}}$ is the minimum virial mass of haloes that host ionizing sources, $\delta_{R,\,n}^{\rm evol}$ is the local, evolved overdensity that is smoothed over a sphere with the radius $R$ and the center at the cell $n$, and $f_{\rm coll}(\delta_{R,\,n}^{\rm evol},\, M_{\rm vir}) $ is the collapsed fraction smoothed over that sphere. The smoothing scale $R$ proceeds from the large to small radius until the above condition is satisfied. If this does not happen with $R$ down to the cell size, then the cell at $n$ is considered as partially ionized with the ionized fraction of $\zeta f_{\rm coll}(\delta_{R,\,n}^{\rm evol},\, M_{\rm vir})$. Without loss of clarity, we use $R$ to denote the scale of cell size\footnote{Strictly speaking, $R$ is the radius of the sphere with the same volume as the cubic cell, i.e.\ $(4\pi/3)\, R^3 = d^3$, or $R= 0.62\,d$, where $d$ is the cell size.} in the rest of this paper. In other words, the neutral fraction field is given by 
\begin{equation}
\label{neutral fraction}
    x_{{\rm HI},\,n} = 
    \begin{cases}
    0\,, &\text{ionized regions;}\\
    1-\zeta f_{\rm coll}(\delta_{R,\,n}^{\rm evol},\, M_{\rm vir})\,, & \text{otherwise}.
    \end{cases} 
\end{equation}

For the CO(1-0) line, we assume that the CO brightness temperature is proportional to the local star formation rate density that is proportional to the collapse fraction on the scale of cell size, $R$. 
\begin{equation}
    T^{\rm{mod,\,coev}}_{{\rm{CO}},\,n} = c_{\rm{CO}}f_{\rm coll}(\delta_{R,\,n}^{\rm evol},\, M_{\rm vir})(1+\delta_{R,\,n}^{\rm evol})\,,
    \label{constTCO}
\end{equation}
where $ c_{\rm CO} = 59.4\,(1+z)^{1/2}$ in units of microkelvins. 

\subsection{Analytical Expression for the Gradient with respect to the Evolved Density}
\label{subsec:analytical-2}

This subsection focuses on deriving an analytical expression for the gradient of the cost function with respect to the evolved density field, $\partial \phi / \partial \delta^{\rm{evol}}_m$, based on the ESMR. It can be written as 
\begin{equation}
\frac{\partial \phi}{\partial \delta_m^{\rm evol}} = \sum_{j,\,\alpha}\frac{\partial \phi}{\partial T_{j,\,\alpha}^{\rm mod}}\frac{\partial T_{j,\,\alpha}^{\rm mod}}{\partial \delta_m^{\rm evol}}\,,
\label{likelihood derivative2}
\end{equation}
where 
\begin{eqnarray}
    \frac{\partial \phi}{\partial T_{j,\,\alpha}^{\rm mod}} &=& C_{\rm cost}\frac{(T^{\rm{mod}}_{j,\,\alpha}-T^{\rm inp}_{j,\,\alpha}) }{(\sigma^{\rm N}_{j})^2}\,\,w_{j} \,, 
      \label{likelihood derivative3} \\
   \frac{\partial T_{j,\,\alpha}^{\rm mod}}{\partial \delta_m^{\rm evol}} &=& C_{\rm trans} \sum_{n\in \alpha} \frac{\partial T^{\rm{mod,\,coev}}_{j,\,n}}{\partial \delta_m^{\rm evol}}.
\end{eqnarray}

The overall normalization factor $C_{\rm cost}$ is a free parameter, so we set $C_{\rm trans}\,C_{\rm cost}=1$ for convenience. 

For the 21~cm map, the derivative is 
\begin{equation}
\label{temperature derivative start}
    \frac{\partial T^{\rm{mod,\,coev}}_{{\rm 21cm},\,n}}{\partial \delta_m^{\rm evol}} = c_{{\rm 21cm}}\left[\delta^K_{n\,m}\,x_{{\rm HI},\,n}+\frac{\partial x_{{\rm HI},\,n}}{\partial \delta_m^{\rm evol}}(1+\delta_{n}^{\rm evol})\right]\,,
\end{equation}
where $\delta^K_{n\,m}$ is the Kronecker delta function. The derivative of the neutral fraction field is $\partial x_{{\rm HI},\,n}/\partial \delta_m^{\rm evol} = 0 $ in ionized regions, and otherwise 
\begin{equation}
    \frac{\partial x_{{\rm HI},\,n}}{\partial \delta_m^{\rm evol}} =  -\zeta\frac{\partial f_{\rm coll}(\delta_{R,\,n}^{\rm evol},\, M_{\rm vir})}{\partial \delta_{R,\,n}^{\rm evol}}\frac{\partial \delta_{R,\,n}^{\rm evol}}{\partial \delta_m^{\rm evol}}\,.
\label{eq:xHI_deriv}
\end{equation}

For the CO map, the derivative is 
\begin{eqnarray}
    \label{temperature derivative mid}
    \frac{\partial T^{\rm{mod,\,coev}}_{{\rm{CO}},\,n}}{\partial \delta_m^{\rm evol}} &=& c_{{\rm CO}}\frac{\partial \delta_{R,\,n}^{\rm evol}}{\partial \delta_m^{\rm evol}}\Bigl [f_{\rm coll}(\delta_{R,\,n}^{\rm evol},\, M_{\rm vir}) \nonumber\\\label{temperature derivative end}
    && +\frac{\partial f_{\rm coll}(\delta_{R,\,n}^{\rm evol},\, M_{\rm vir})}{\partial \delta_{R,\,n}^{\rm evol}}(1+\delta_{R,\,n}^{\rm evol})\Bigr ]\,.
\end{eqnarray}

Both 21~cm and CO maps are determined by the local collapse fraction $f_{\rm coll}$ (see Equations~\ref{neutral fraction}, \ref{constTCO}, \ref{eq:xHI_deriv}, \ref{temperature derivative mid}). We employ the Press-Schecter (PS) halo mass function \citep{1974ApJ...187..425P} to calculate the collapse fraction and its derivative, and normalize them with the Sheth-Torman (ST) correction \citep{1999MNRAS.308..119S}. 
\begin{equation}
    f_{\rm coll}(\delta_{R,\,n}^{\rm evol},\, M_{\rm vir}) = \frac{\bar{f}_{\rm ST}}{\bar{f}_{\rm PS}}\,{\rm erfc}\left[\frac{\delta_c(z)-\delta_{R,\,n}^{\rm evol}}{\sqrt{2(\sigma^2_{\rm min}-\sigma^2_{R}})}\right]\,,
\end{equation}
where $\bar{f}_{\rm ST}$ and $\bar{f}_{\rm PS}$ are the mean ST and PS collapse fractions, respectively. $\delta_c(z)$ is the critical overdensity for collapse at redshift $z$. $\sigma_R$ is the variance of density fluctuations at the smoothing scale $R$, and $\sigma_{\rm min}=\sigma(M_{\rm vir})$ is the variance at the scale corresponding to the minimum halo mass $M_{\rm vir}$. 

The last ingredient is the derivative $\partial \delta_{R,\,n}^{\rm evol}/\partial \delta_m^{\rm evol}$ that is involved in the calculations for both derivatives of the 21~cm and CO maps (see Equations~\ref{eq:xHI_deriv} and \ref{temperature derivative mid}). The smoothed overdensity $\delta_{R,\,n}^{\rm evol}$ is the convolution of the overdensity $\delta_n^{\rm evol}$ and a window function. If we rewrite it in the form of matrix multiplication as $\delta_{R,\,n}^{\rm evol} = \sum_m \textbf{F}_{n\,m}\,\delta_m^{\rm evol}$, then $\partial \delta_{R,\,n}^{\rm evol}/\partial \delta_m^{\rm evol} = \textbf{F}_{n\,m}$, where the matrix $\textbf{F}_{n\,m}$ is given by the smoothing kernel and the smoothing scale $R$. In practice, we exploit the symmetry $\textbf{F}_{n\,m} = \textbf{F}_{m\,n}$ to simplify the calculation when summations over $n$ and $\alpha$ are put together in Equation~(\ref{likelihood derivative2}). 

\begin{figure}
    \centering
    \includegraphics[width=0.9\columnwidth]{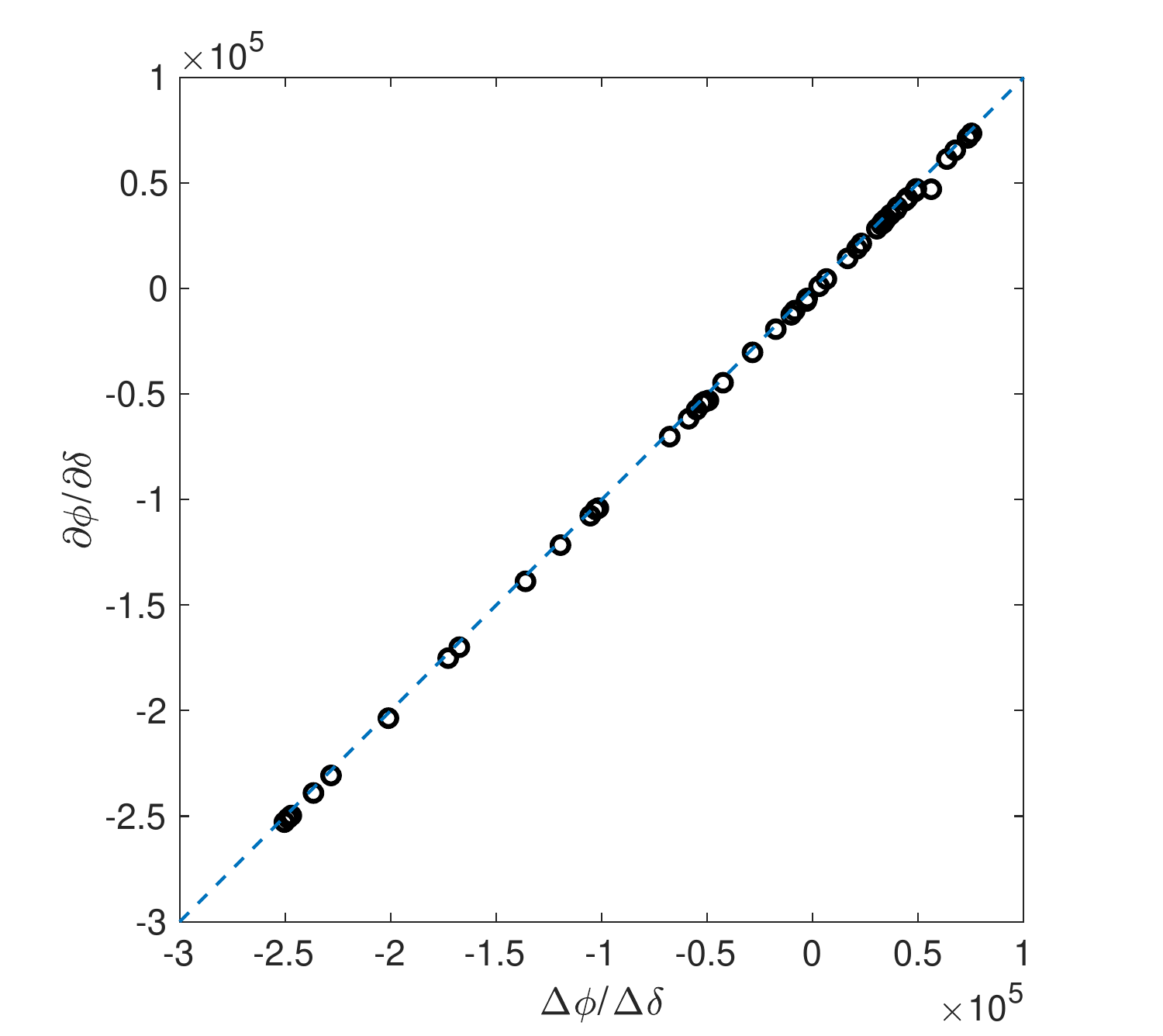}
     \caption{Accuracy test of the analytical formalism for $\partial \phi/\partial \boldsymbol{\delta}^{\rm ini}$. We compare the gradients calculated analytically (``$\partial\phi/\partial\delta$'') and numerically(``$\Delta\phi/\Delta\delta$'') at 50 random locations, respectively. The diagonal dashed line indicates the perfect matching.}
    \label{fig:gradient_test}
\end{figure}

Now we can calculate the derivative $\partial \phi/\partial \boldsymbol{\delta}^{\rm ini}$ for the 21~cm and CO intensity maps with an analytical formalism. To test the accuracy of this formalism, we choose 50 locations randomly and compare the derivatives $\partial \phi / \partial \delta^{\rm{ini}}_n $ that are calculated analytically and numerically in Figure~\ref{fig:gradient_test}, respectively. For the numerical implementation, we shift the initial density at the chosen location (with the cell index $n$) with a small variation $\Delta\delta^{\rm ini}_n=10^{-4}\delta^{\rm ini}_n$ and calculate the resulting difference of the cost function $\Delta\phi$, and then use the finite difference method to estimate the gradient, $\Delta \phi / \Delta \delta^{\rm{ini}}_n$. The comparison shows that the analytical results are in good agreement with the numerical results. 

\begin{table*}
    \centering
    \caption{Mock Observations of the 21~cm and CO Brightness Temperature Fields --- the Redshift, the Mean Neutral Faction, the Global Average, and the Standard Deviation of the 21cm Field and the CO Field. We Also Show the Goodness of Reconstruction for the Resimulated Fields. } 
    \begin{tabular}{ccccccc}
    \hline \hline 
       $z$ & $\bar{x}_{\rm HI}$ & $\bar{T}_{\rm 21cm}$ [mK] & $\bar{T}_{\rm CO}$ [$\mu$K] &  $\sigma_{\rm 21cm}$ [mK]  & $\sigma_{\rm CO}$ [$\mu$K] & $L_{\rm tot}$ \\
      \hline 
       7.56 & 0.25 & 4.675 & 8.759 & 5.615 & 14.20 & 0.0495  \\
       \hline 
       8.20 & 0.5 & 10.68 & 6.486 & 6.128 & 11.22 & 0.0673 \\
       \hline 
       9.54 & 0.75 & 18.83 & 3.259 & 4.688 & 6.460 & 0.0455 \\
       \hline 
    \end{tabular}    
    \label{table of uncertainty}
\end{table*}

\subsection{Simulations}
\label{subsec:sim}

We perform semi-numerical simulations of reionization with a modified\footnote{The modification is made to simulate the CO line intensity maps. The original version of the {\tt 21cmFAST} code is publicly available at \url{https://github.com/21cmfast/21cmFAST}.} version of the code {\tt 21cmFAST} \citep{Mesinger201121cmfast}.  It is based on the semi-numerical treatment of cosmic reionization with the excursion-set approach \citep{2004ApJ...613....1F} to identify ionized regions. Our simulations were performed on a cubic box of 368 comoving ${\rm Mpc}$ on each side, with $128^3$ grid cells. We choose a reference simulation with the parameter value $\zeta = 25$ and $M_{\mathrm{vir}} = 5 \times 10^{8}\,M_\odot$. Note that these reionization parameters are {\it fixed} in the initial density reconstruction. 

The reconstruction of initial density fields considered in this paper is made in the comoving cubic volume where the mock 21~cm and CO brightness temperature fields are measured at three different redshifts $z=7.56$, $8.20$ and $9.54$ (corresponding to three stages of reionization, $\bar{x}_{\rm HI} = 0.25$, $0.50$ and $0.75$), independently. We set the initial density field of the mock observations at the redshift $z^{\rm ini}=20$. The mock observations are constructed from the reference simulation in a coeval manner in the sense that we neglect the lightcone effect \citep{2012MNRAS.424.1877D,2014MNRAS.442.1491D}  across the simulation box along the line-of-sight. We list the mean 21~cm and CO brightness temperature and their standard deviations in Table~\ref{table of uncertainty}. We also assume the pixel resolution of observations corresponds to $5.75$ comoving Mpc in the coarse-grained cells.

\begin{figure*}
    \centering
    \includegraphics[width= 0.65\columnwidth]{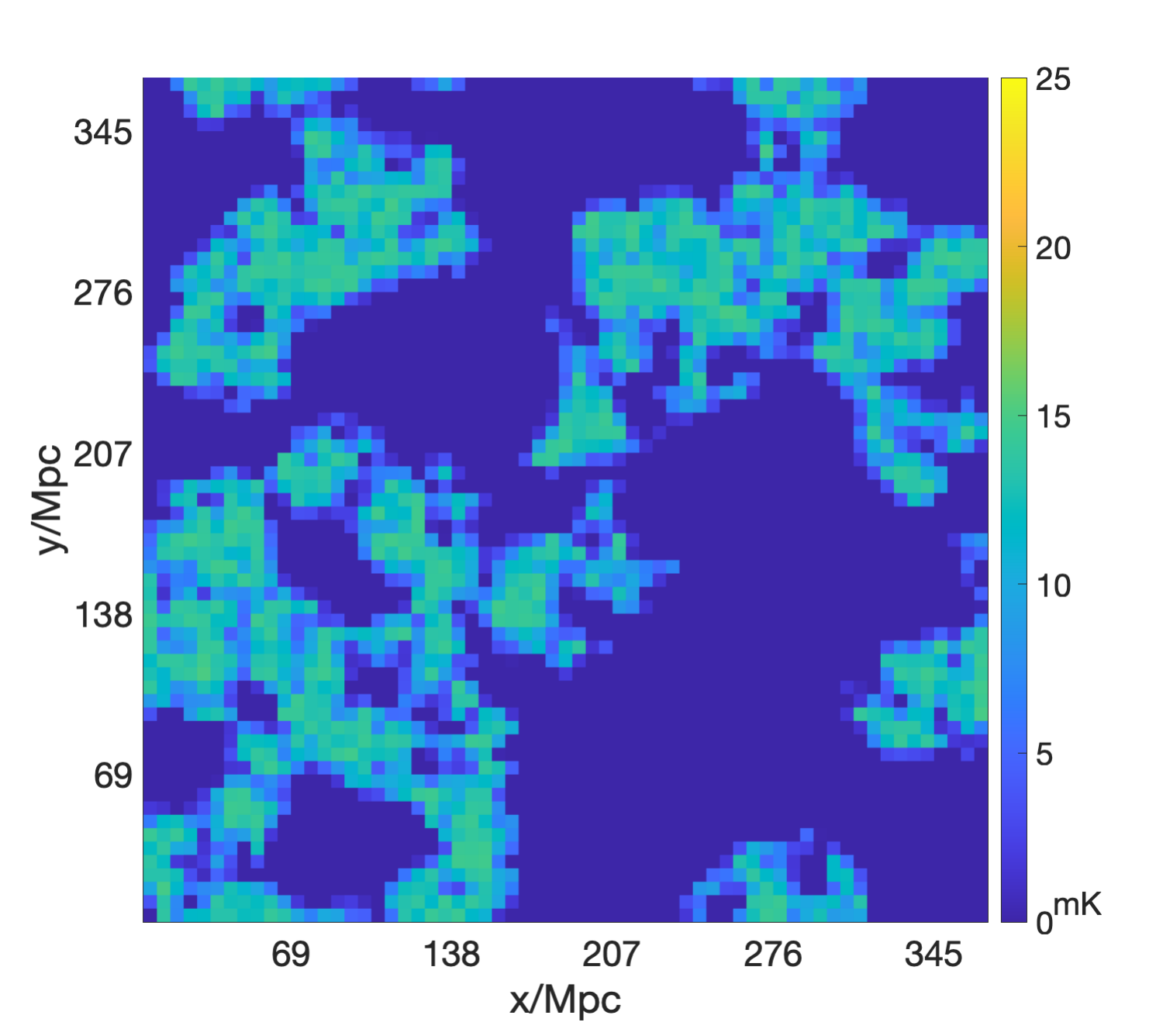}
    \includegraphics[width= 0.65\columnwidth]{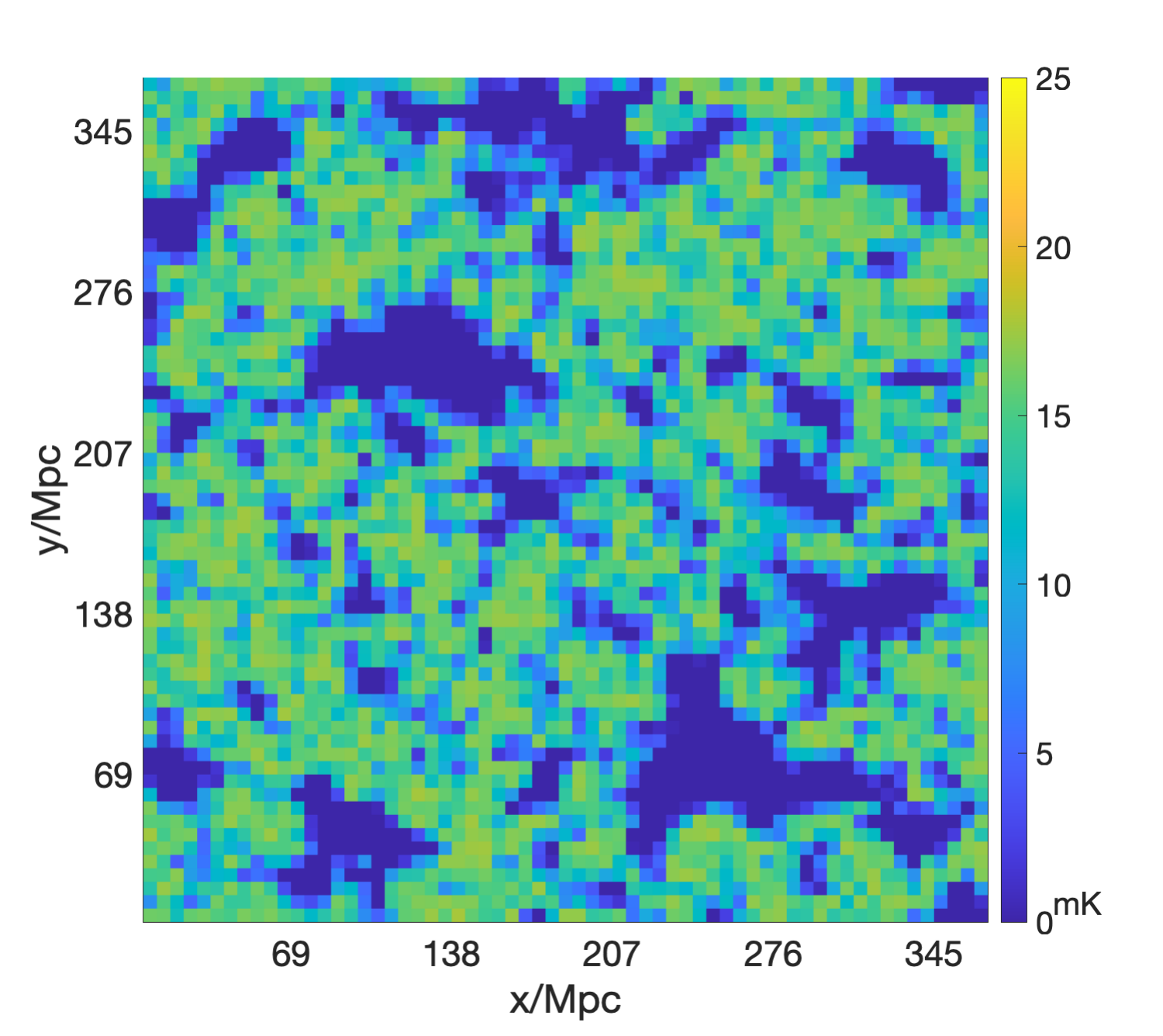}
    \includegraphics[width= 0.65\columnwidth]{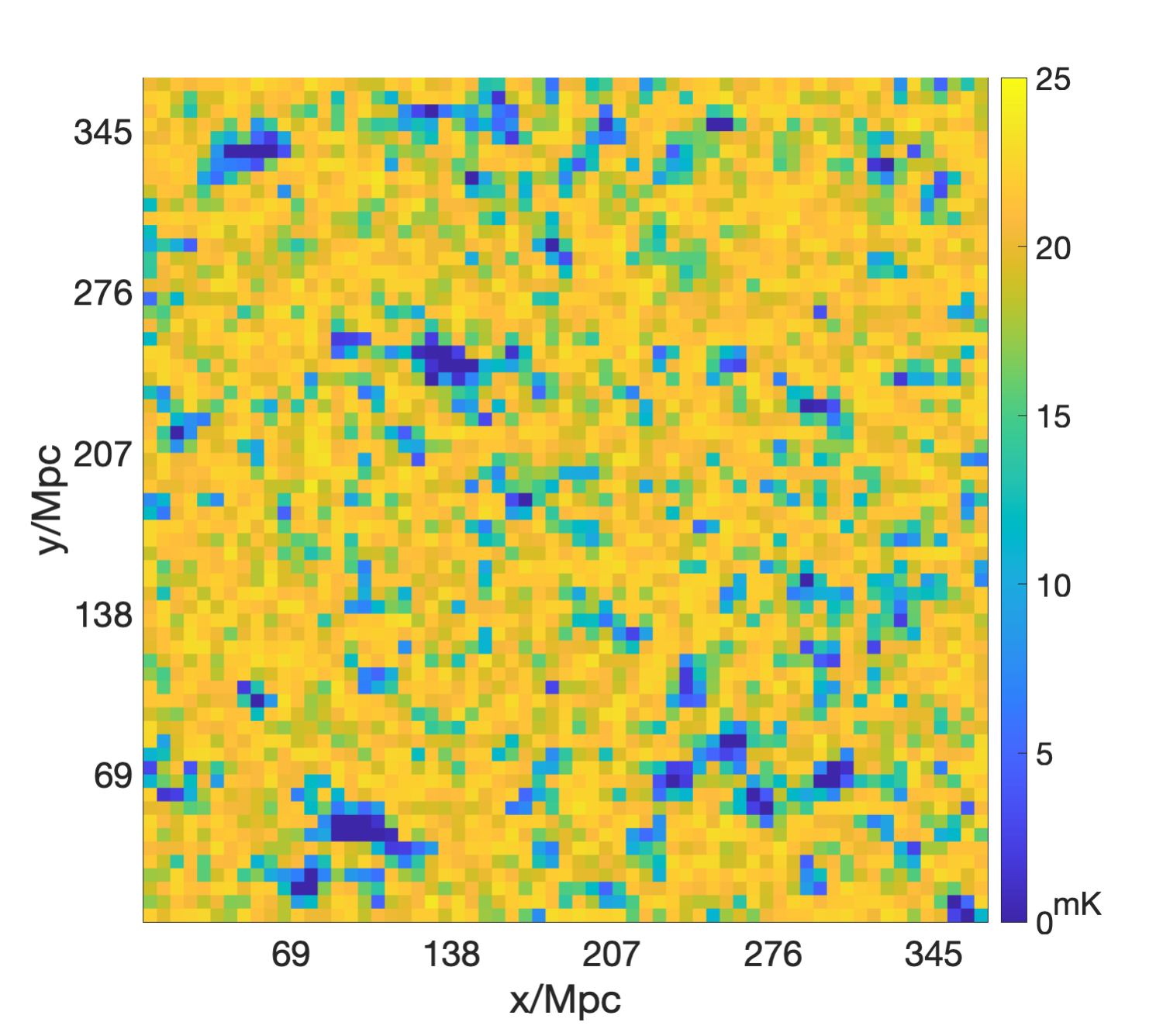}
    \includegraphics[width= 0.65\columnwidth]{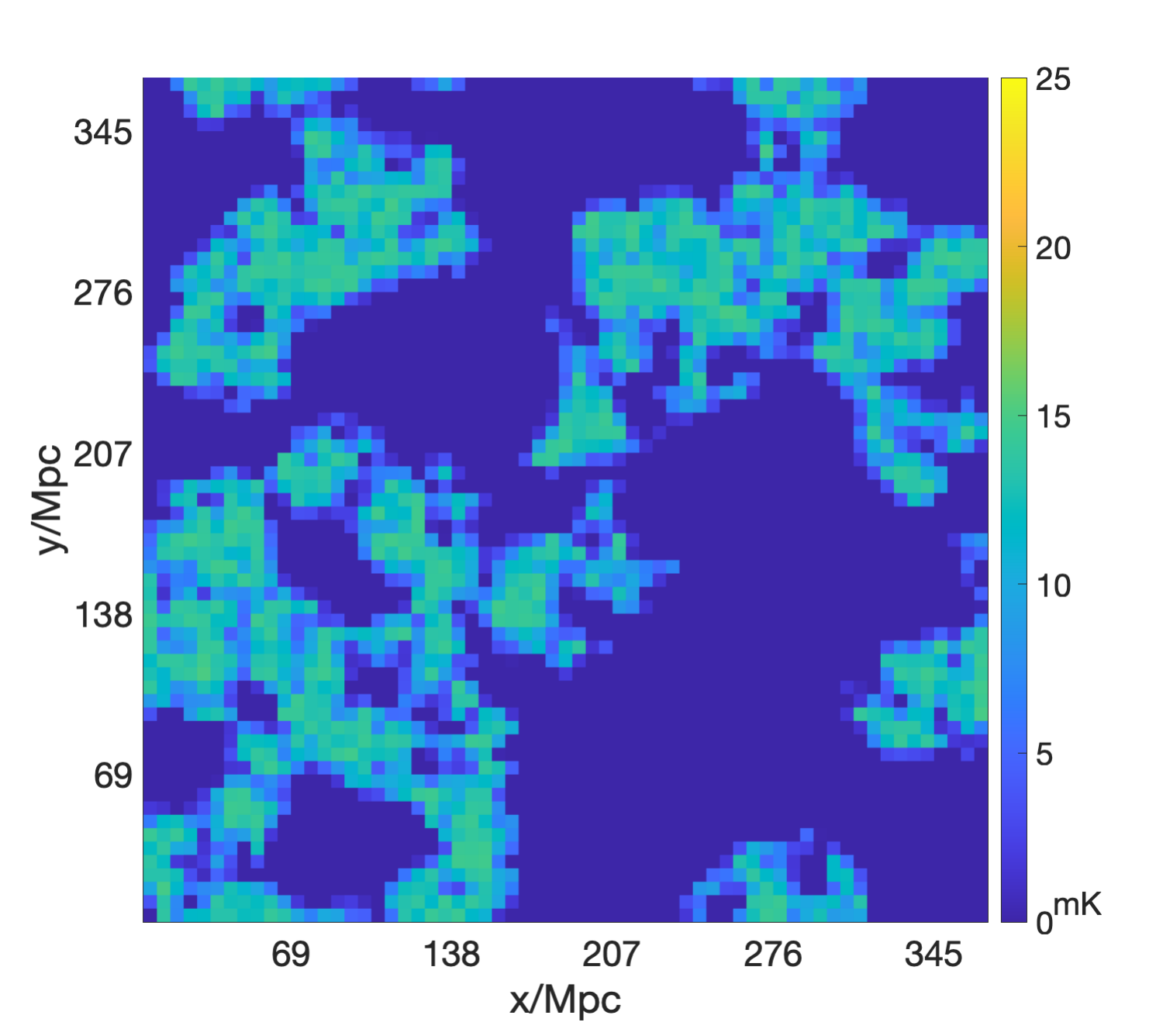}
    \includegraphics[width= 0.65\columnwidth]{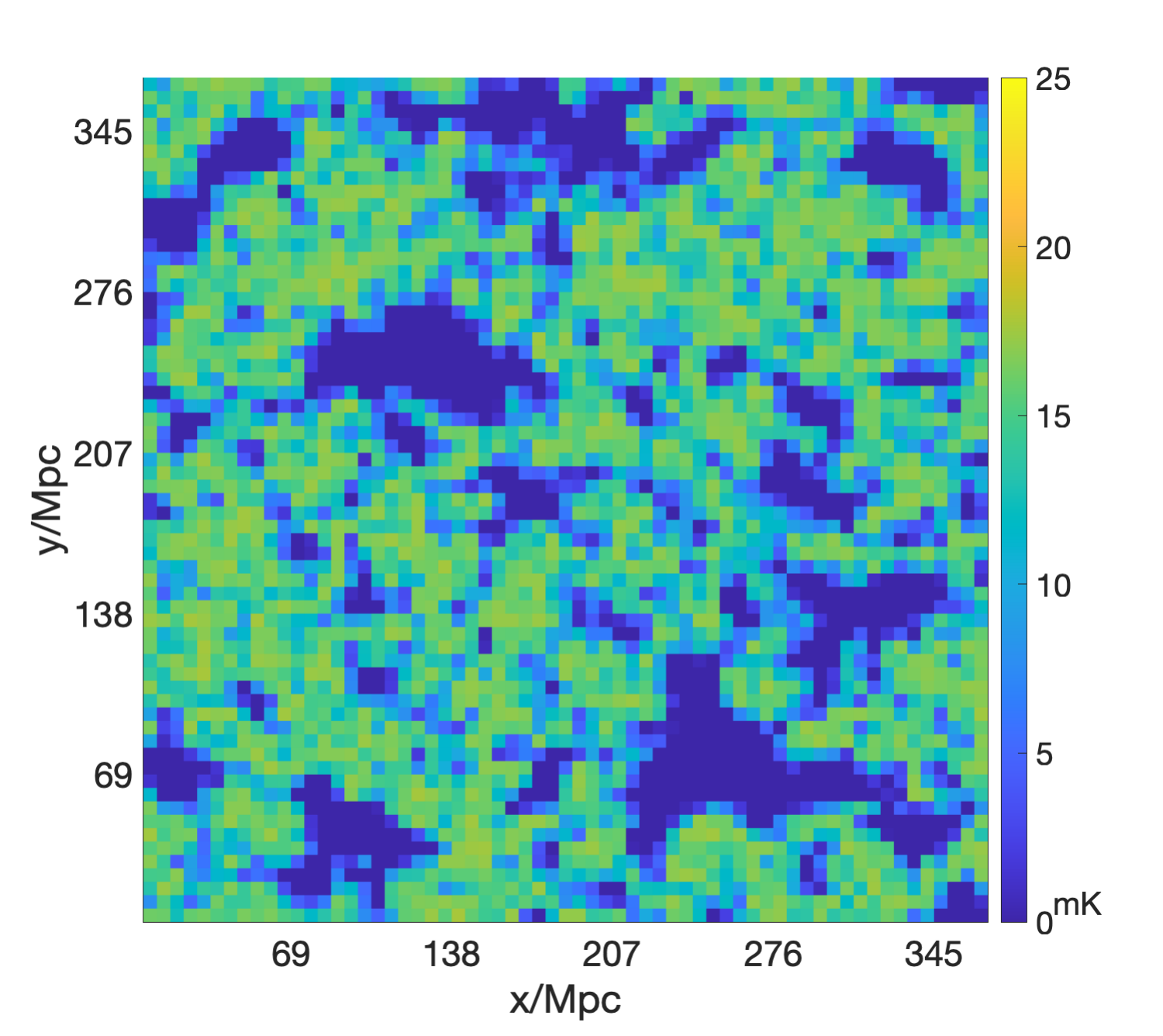}
    \includegraphics[width= 0.65\columnwidth]{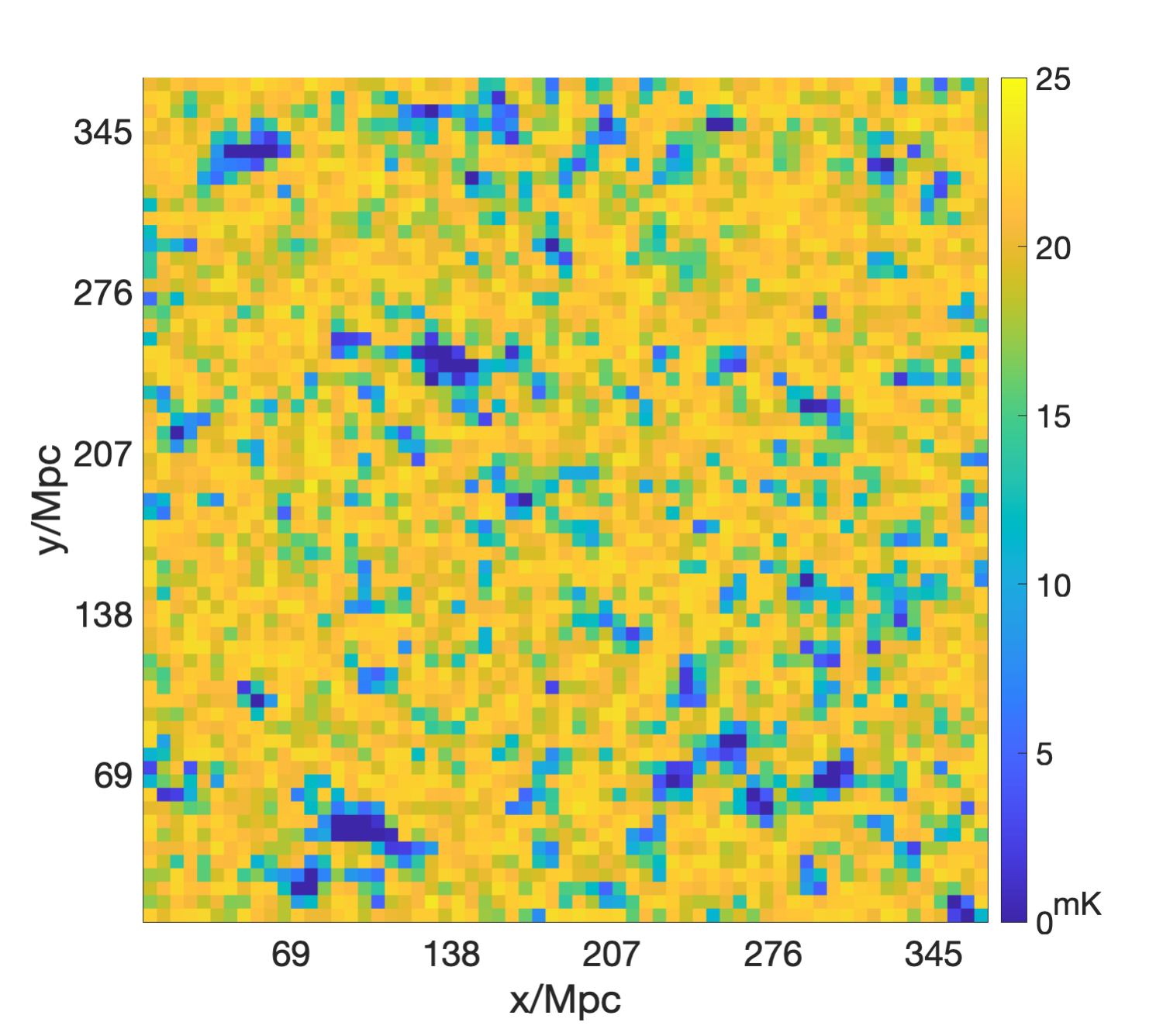}
    \caption{The 21~cm brightness temperature maps of the input field (top) and the resimulated field (bottom) in units of millikelvins. We show the maps in a slice of simulated coeval box with 368 comoving Mpc on each side, (from left to right) at redshift $z=7.56$, $8.20$ and $9.54$, corresponding to global neutral fraction $\bar{x}_{\rm HI} = 0.25$, $0.50$ and $0.75$, respectively.}
    \label{fig:21cm_images}
\end{figure*}

\begin{figure*}
    \centering
    \includegraphics[width= 0.65\columnwidth]{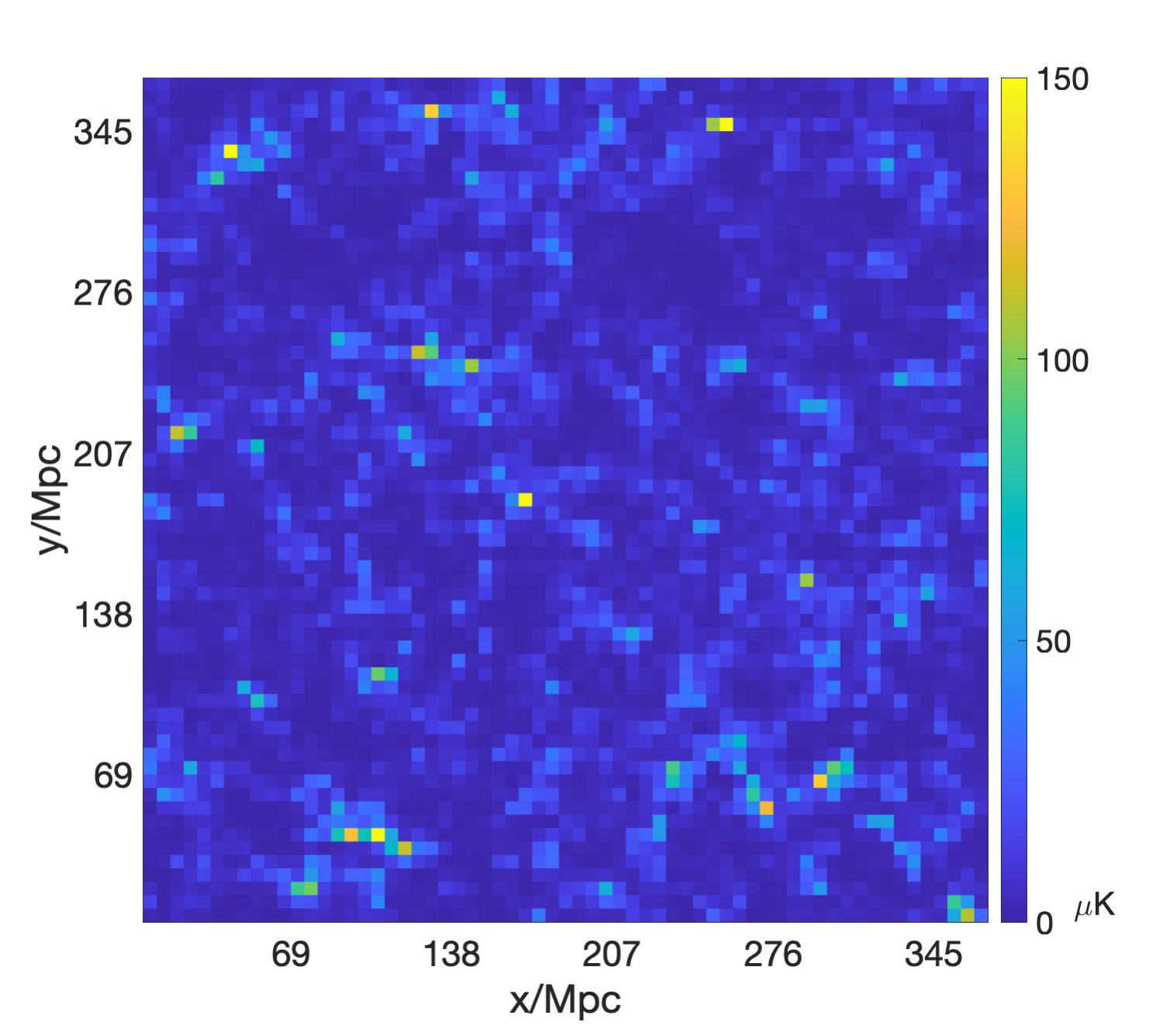}
    \includegraphics[width= 0.65\columnwidth]{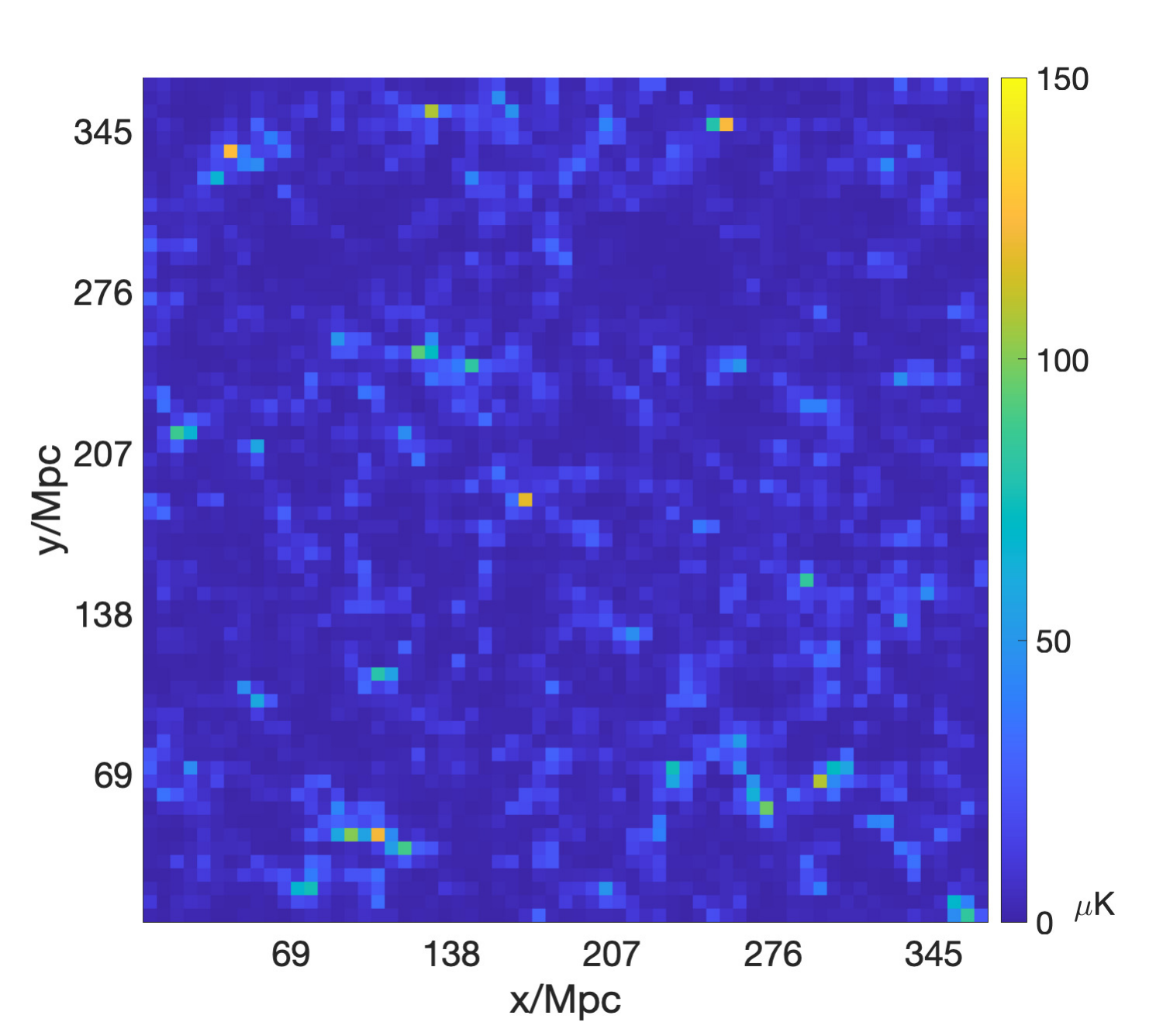}
    \includegraphics[width= 0.65\columnwidth]{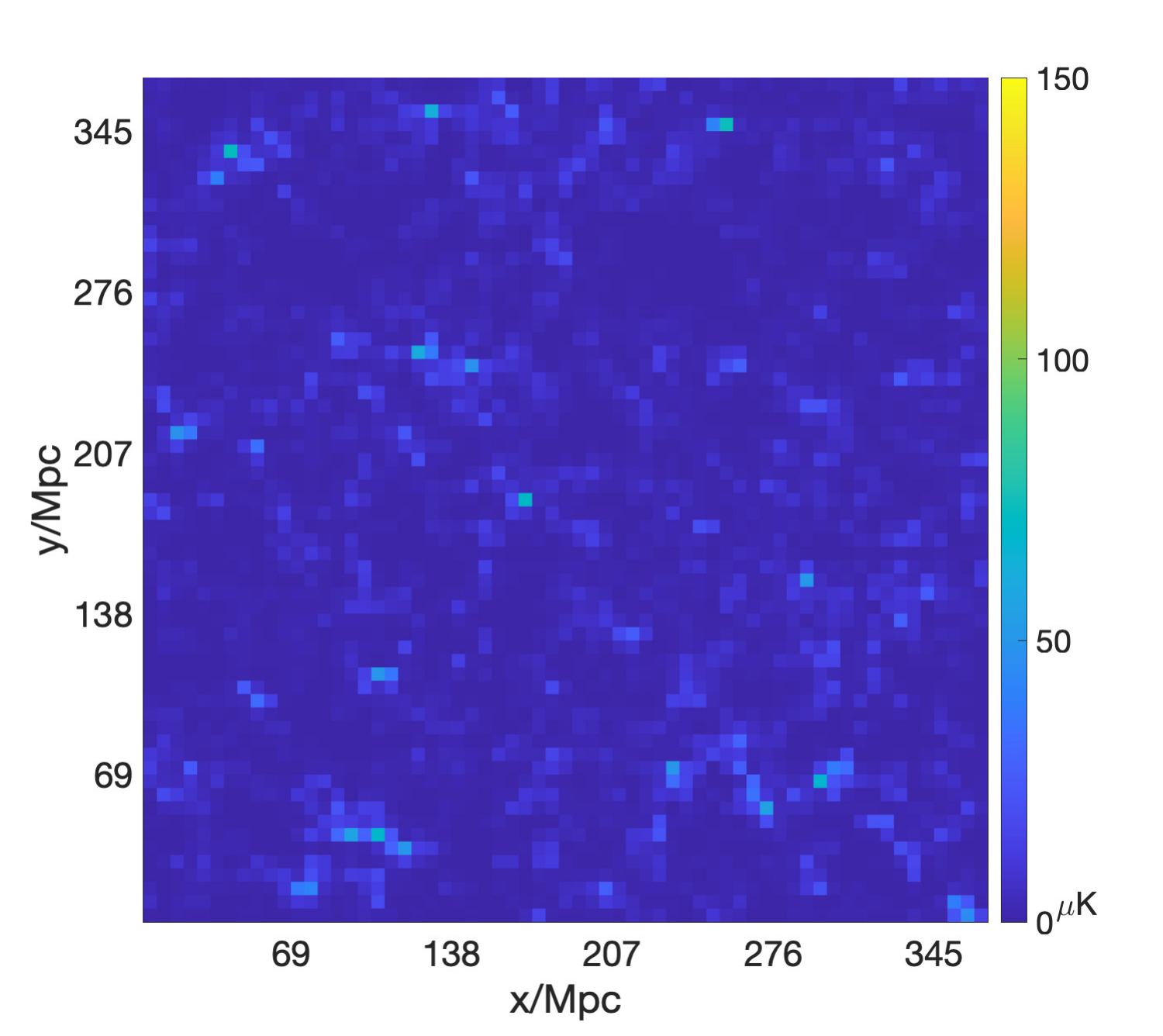}
    \includegraphics[width= 0.65\columnwidth]{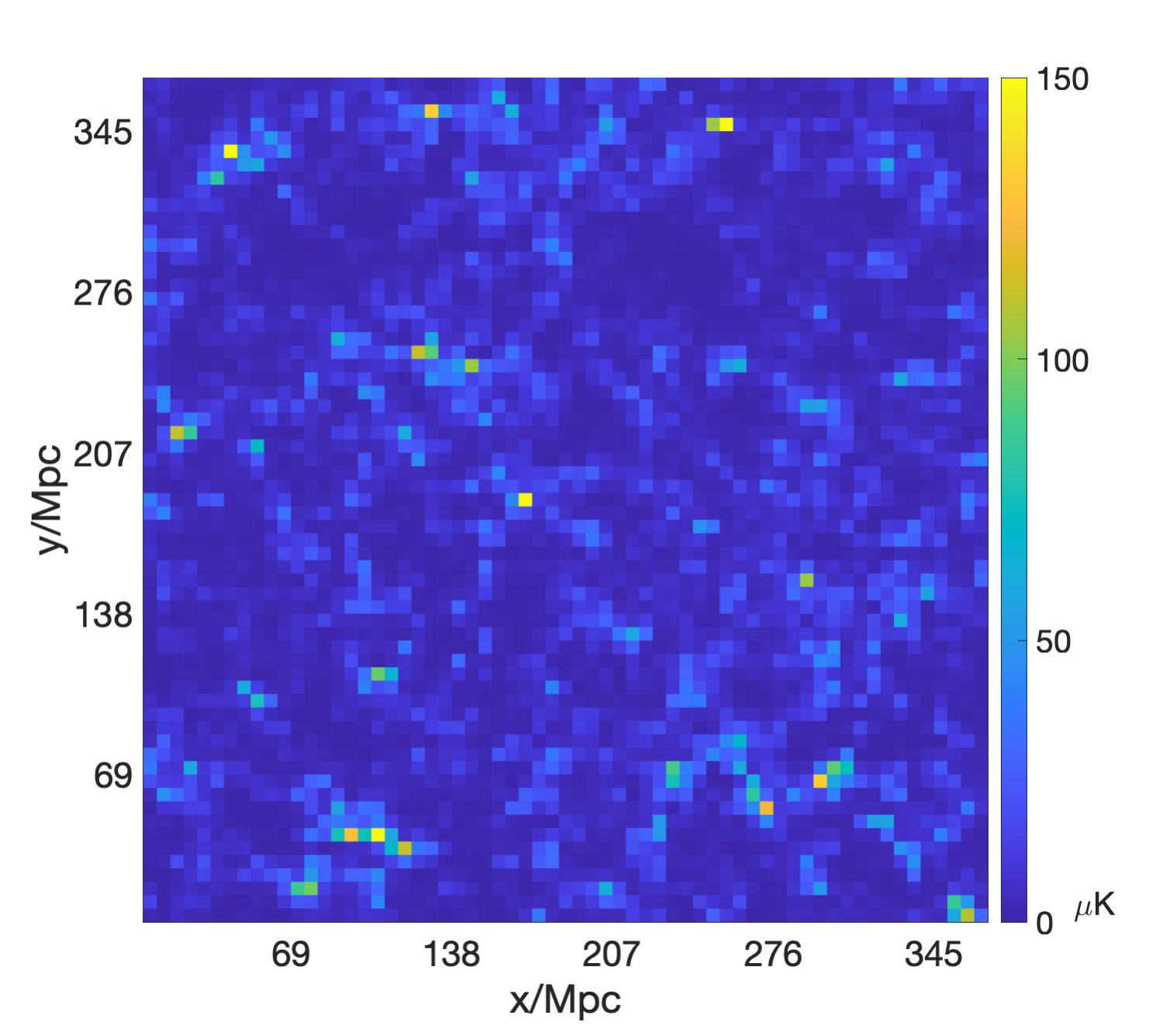}
    \includegraphics[width= 0.65\columnwidth]{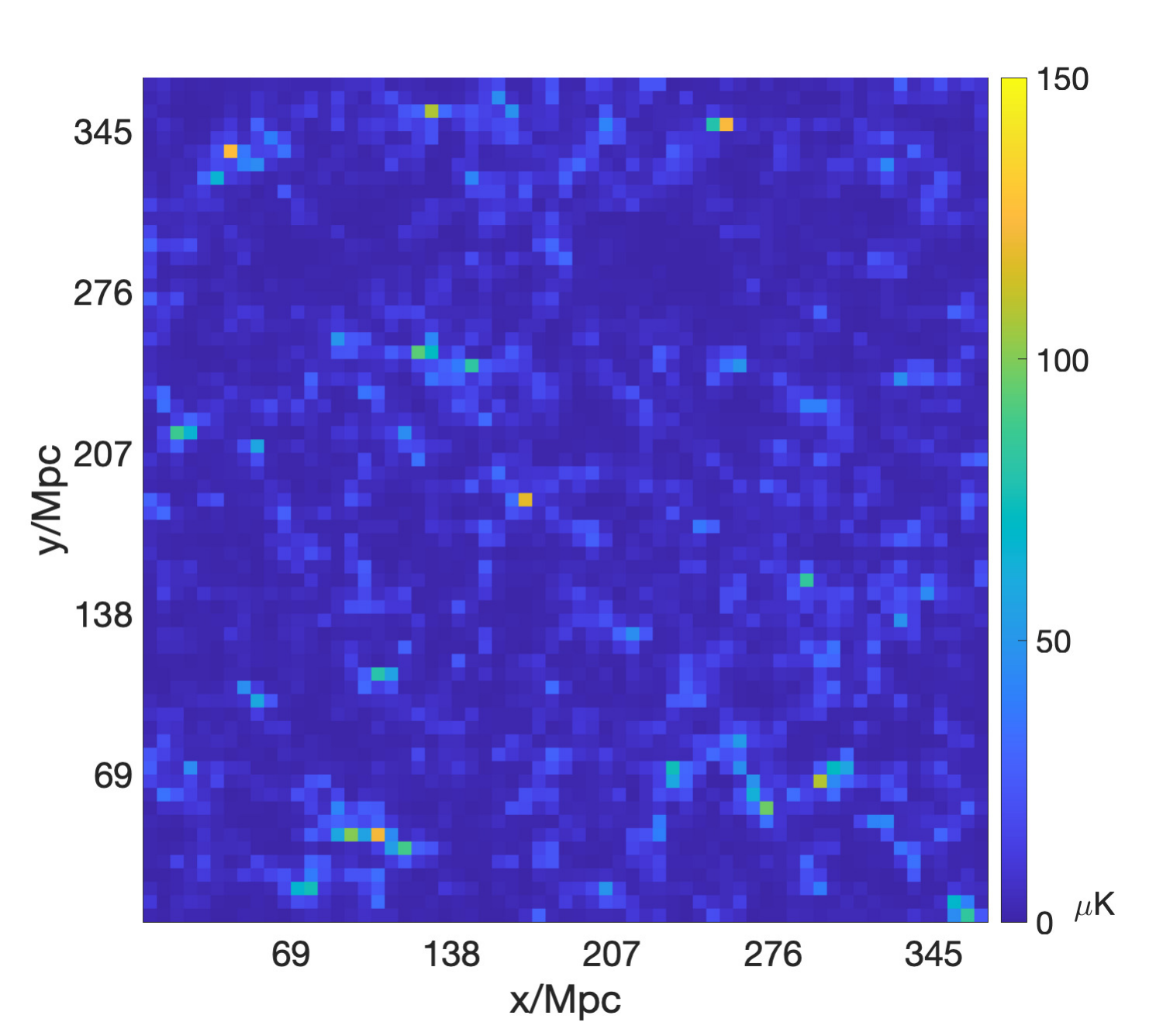}
    \includegraphics[width= 0.65\columnwidth]{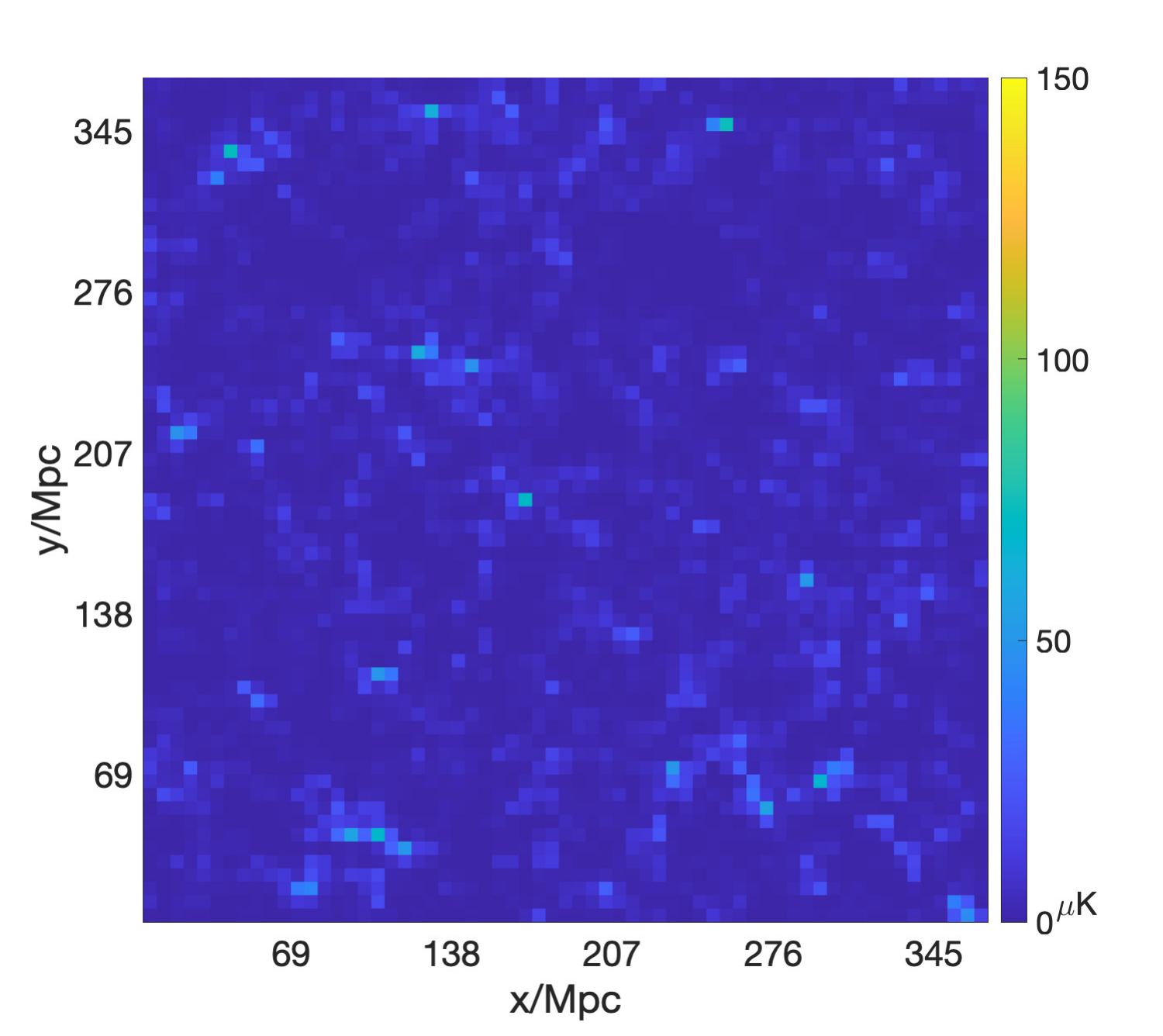}
    \caption{Same as Figure~\ref{fig:21cm_images} but for the CO brightness temperature maps in units of microkelvins.}
    \label{fig:CO_images}
\end{figure*}

\begin{figure*}
    \includegraphics[width= 0.65\columnwidth]{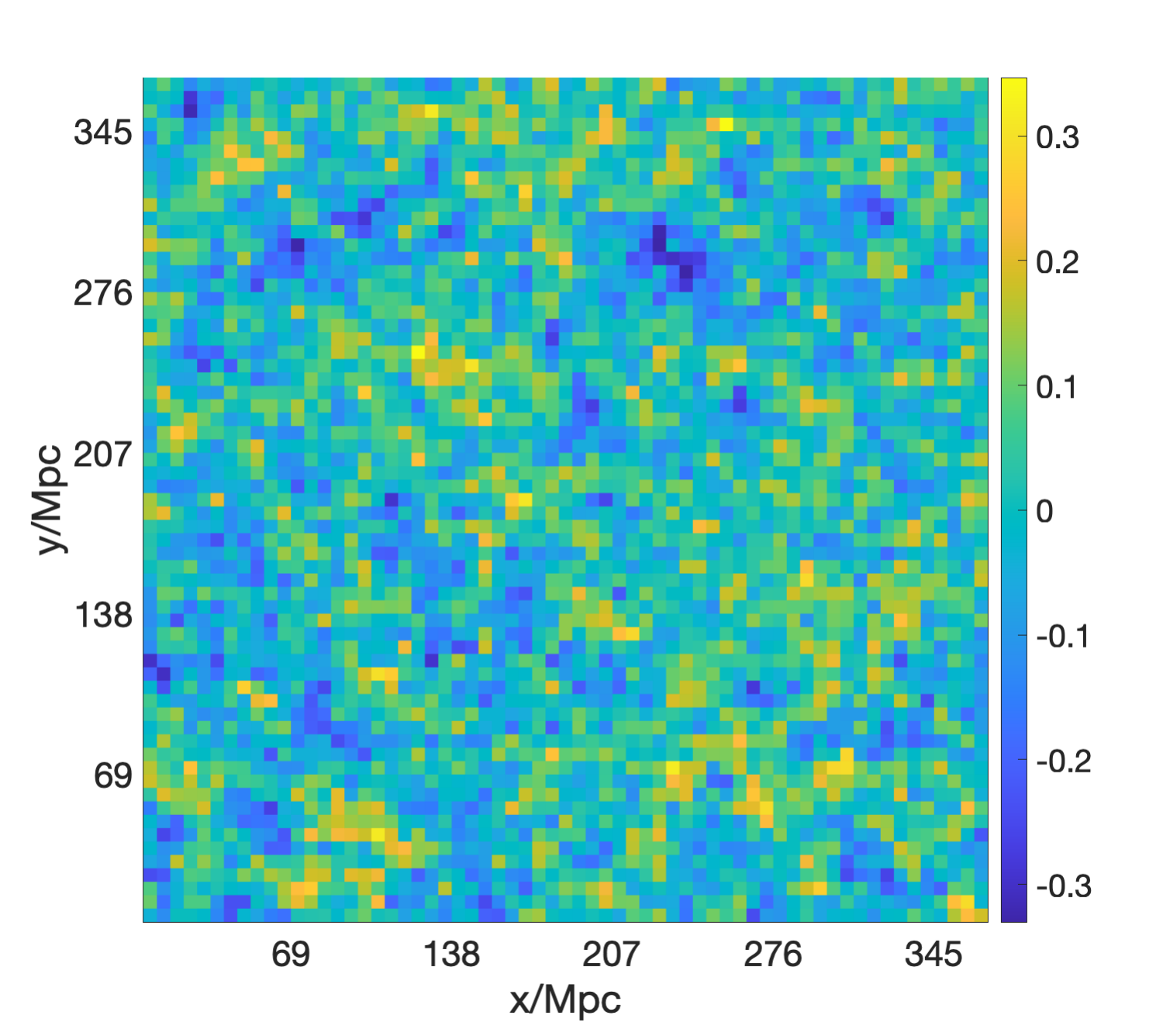}\\
    \includegraphics[width= 0.65\columnwidth]{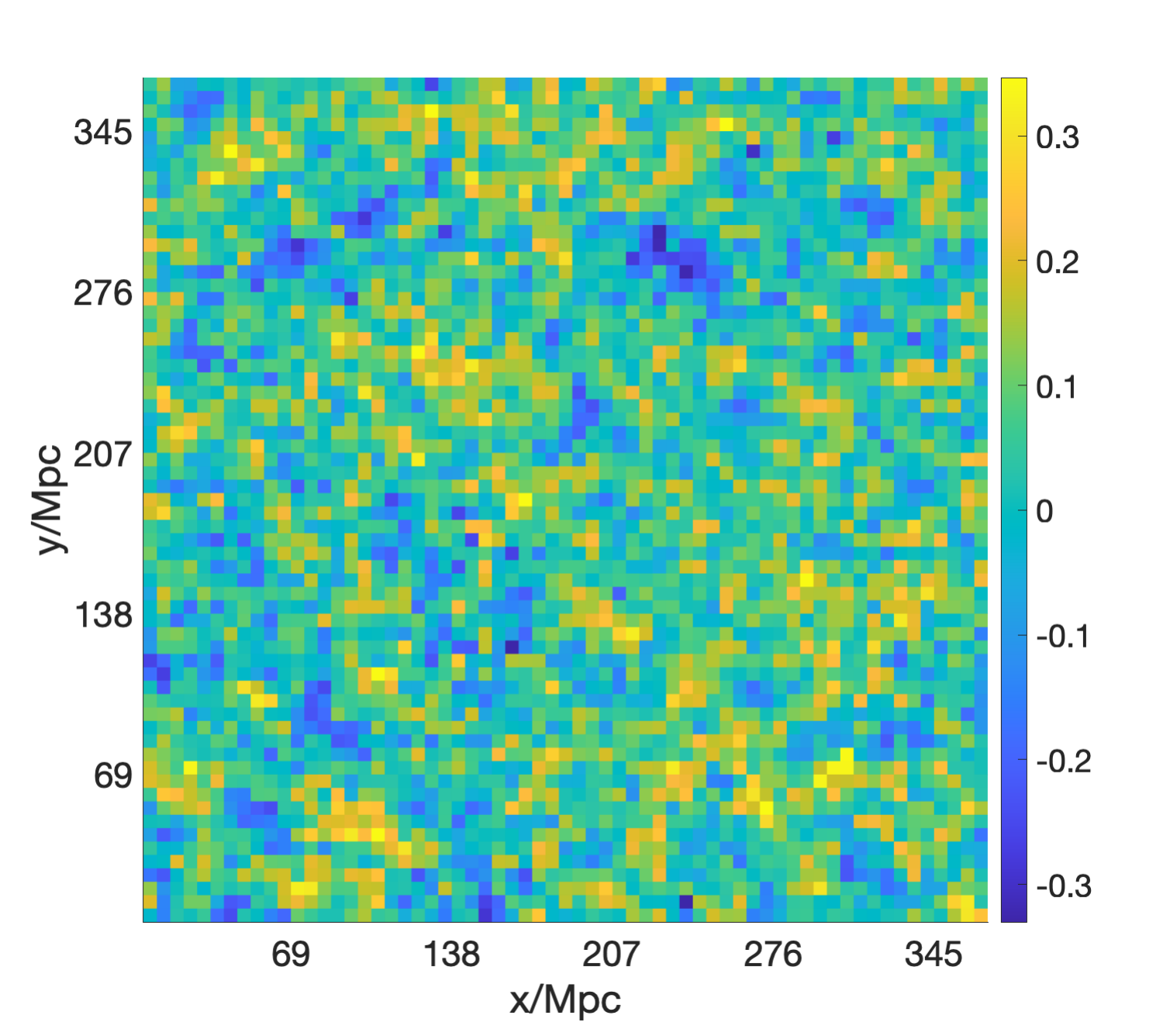}
    \includegraphics[width= 0.65\columnwidth]{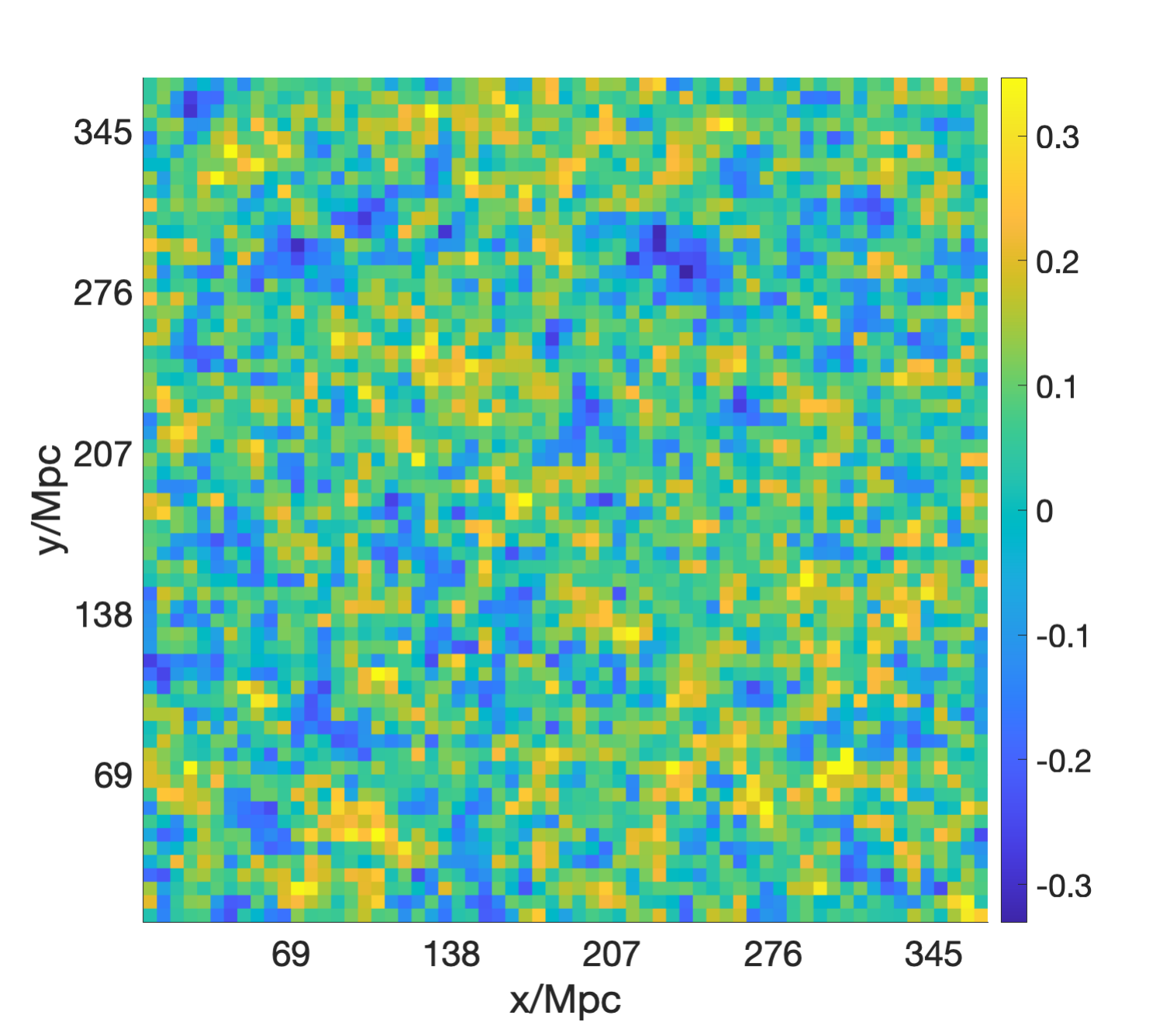}
    \includegraphics[width= 0.65\columnwidth]{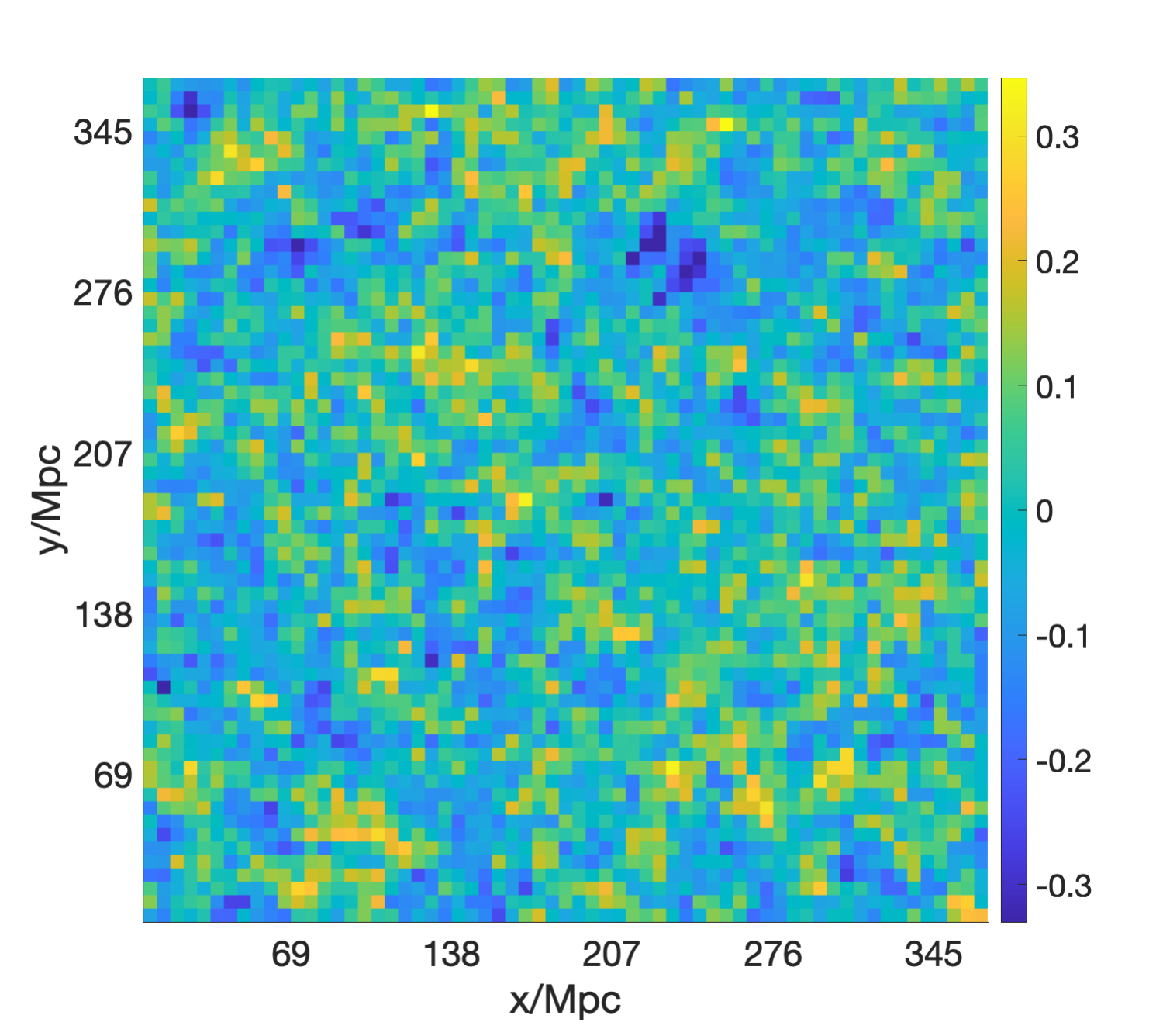}\\
    \includegraphics[width= 0.65\columnwidth]{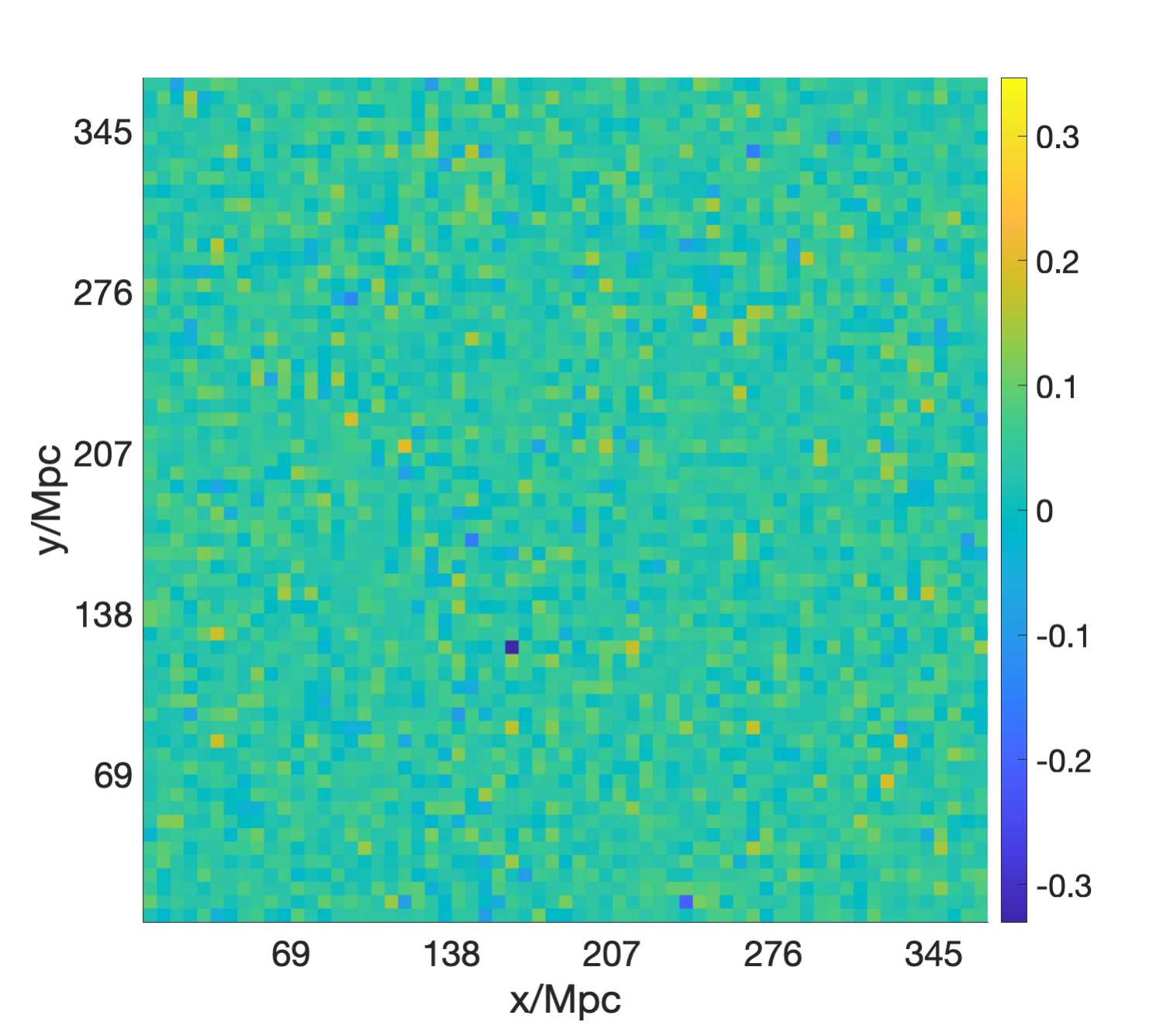}
    \includegraphics[width= 0.65\columnwidth]{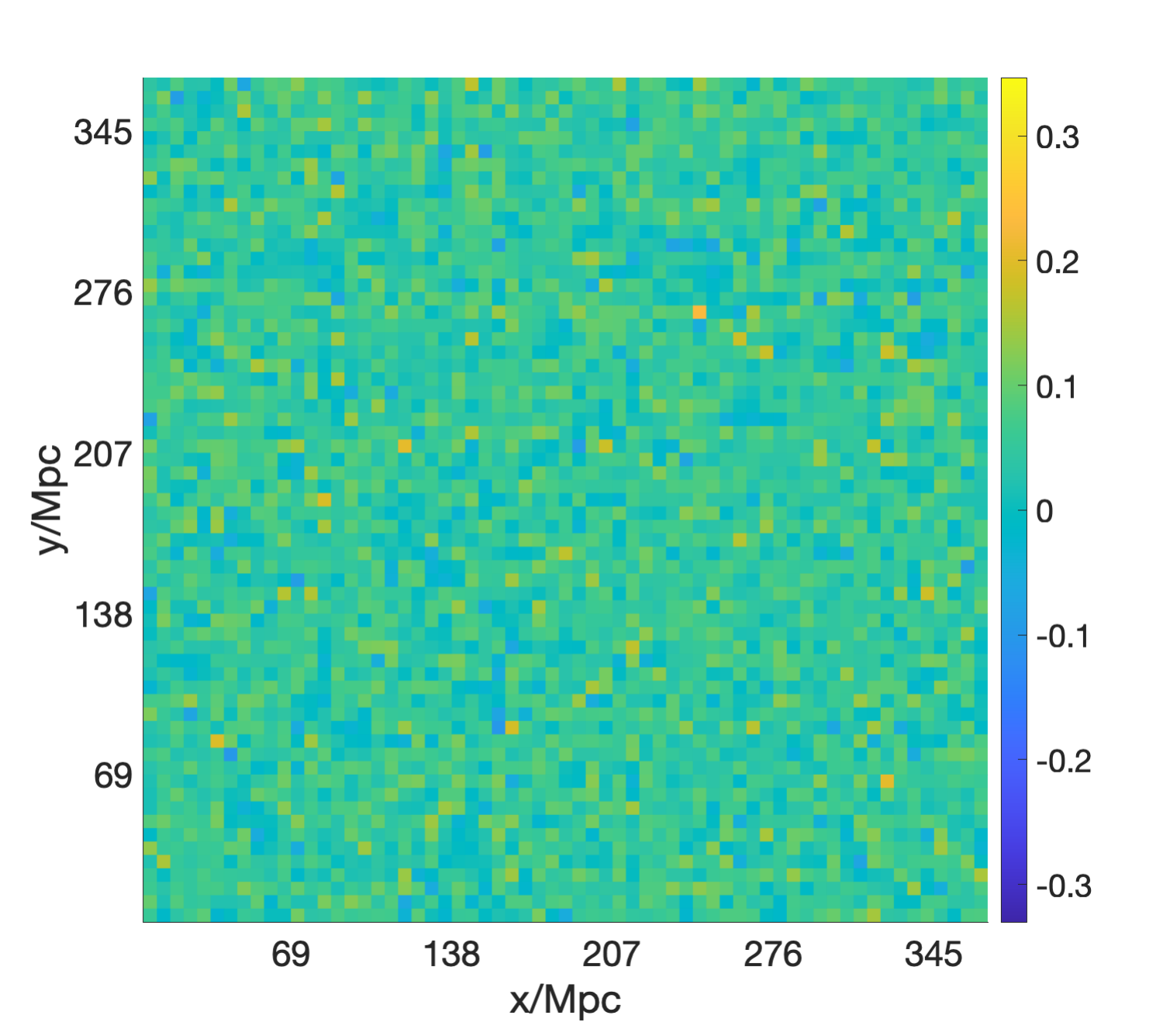}
    \includegraphics[width= 0.65\columnwidth]{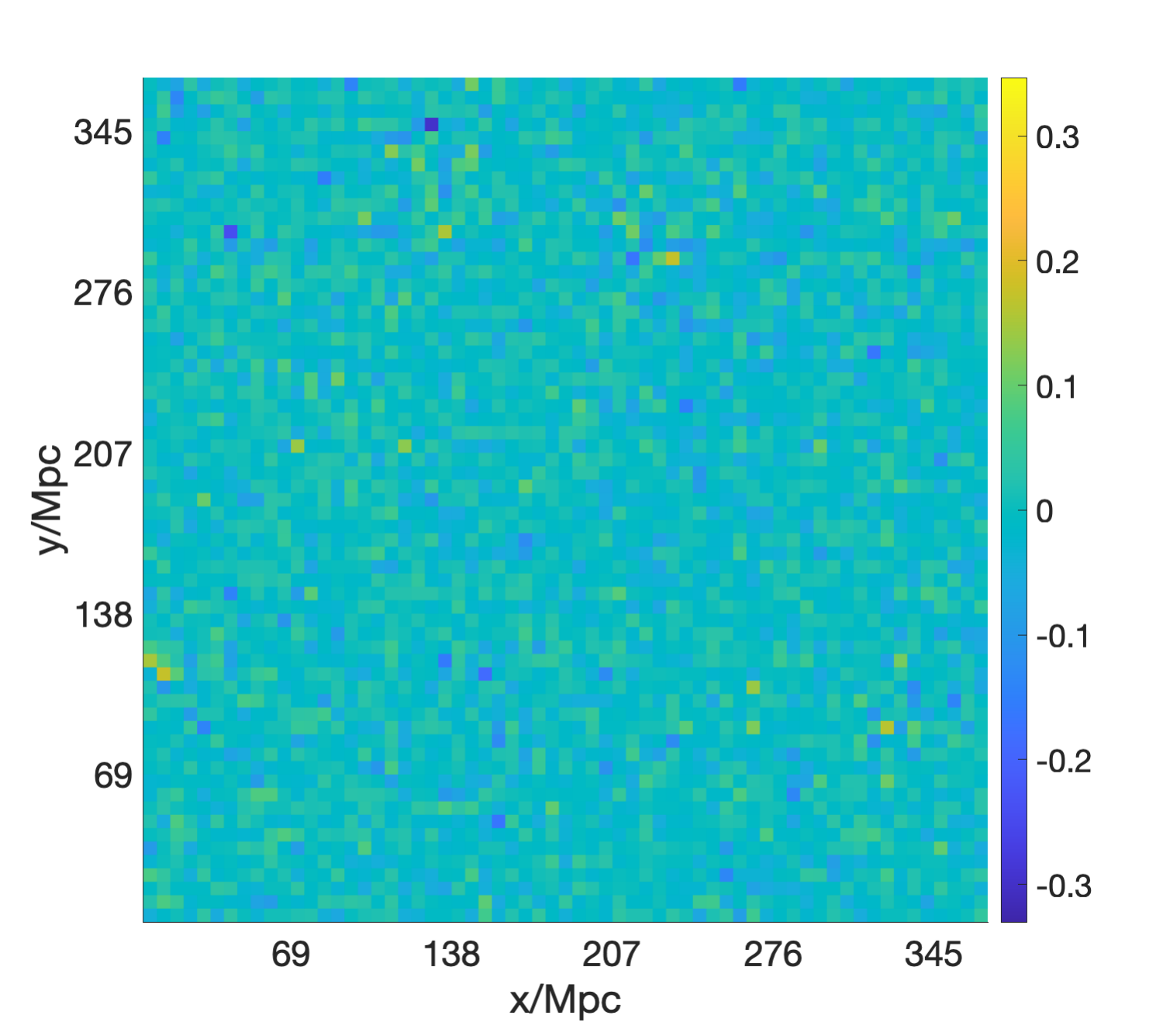}
    \caption{The initial overdensity field $\boldsymbol{\delta}^{\rm ini}$ in a slice of comoving volume with 368 Mpc on each side. We show the input, true field (top left), and the reconstructed field (middle) using the mock observations of the 21~cm and CO maps in a coeval box (from left to right) at redshift $z=7.56$, $8.20$, and $9.54$, respectively. For the purpose of comparison, we also show the residual between the reconstructed and the true initial overdensity (bottom). }
    \label{fig:image_comparison_pure}
\end{figure*}

\begin{figure*}
    \centering
    \includegraphics[width= 0.65\columnwidth]{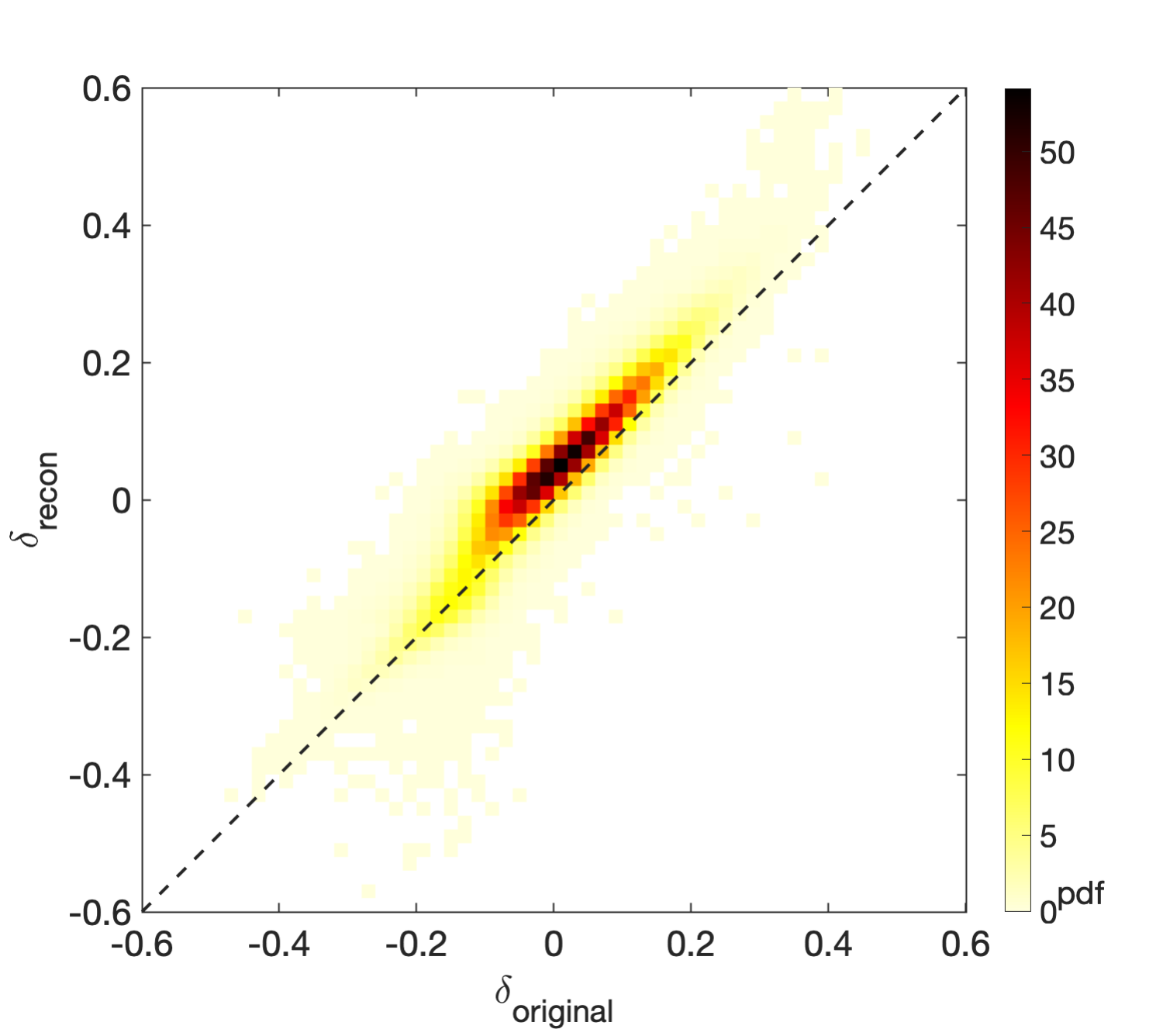}
    \includegraphics[width= 0.65\columnwidth]{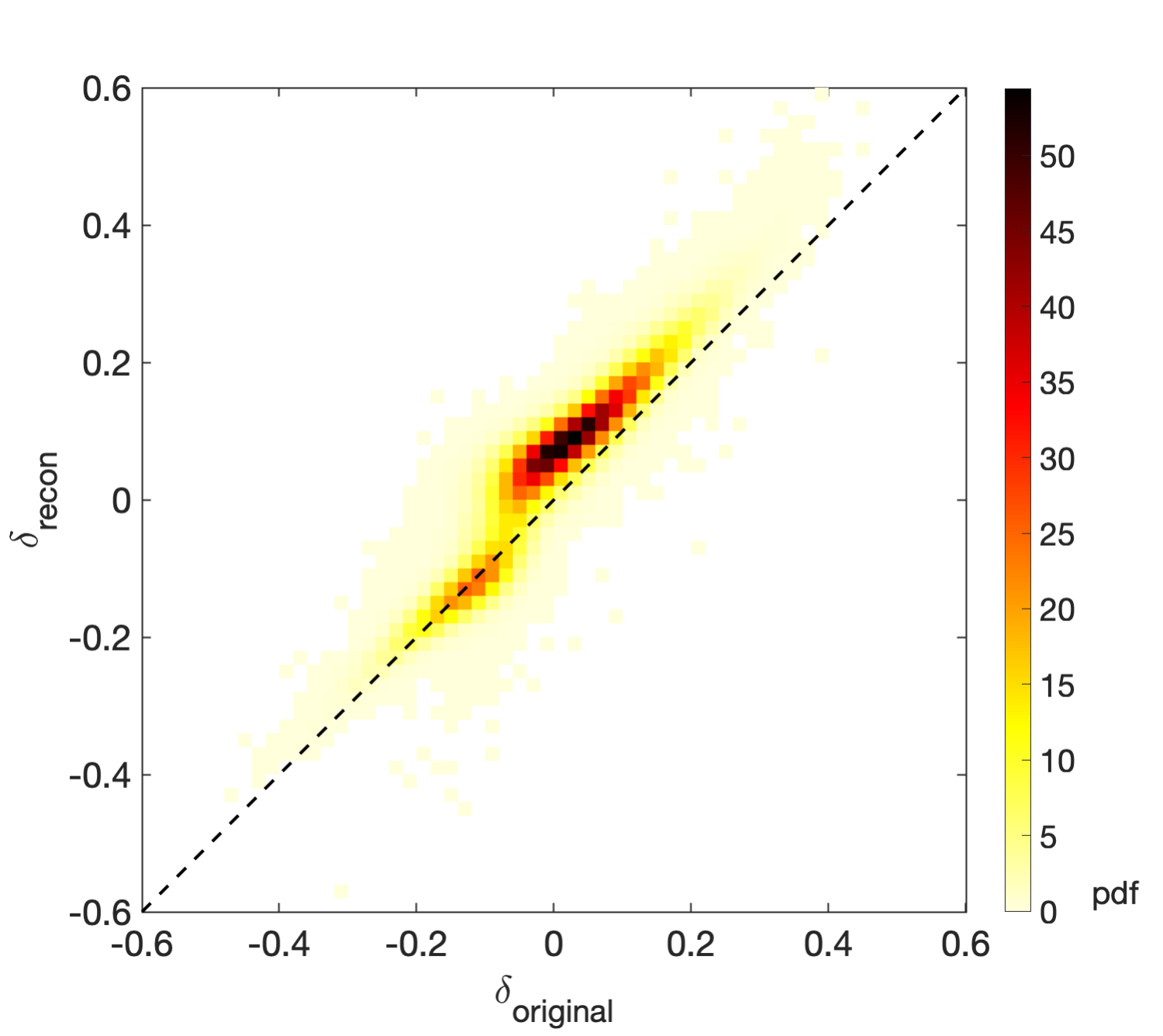}
    \includegraphics[width= 0.65\columnwidth]{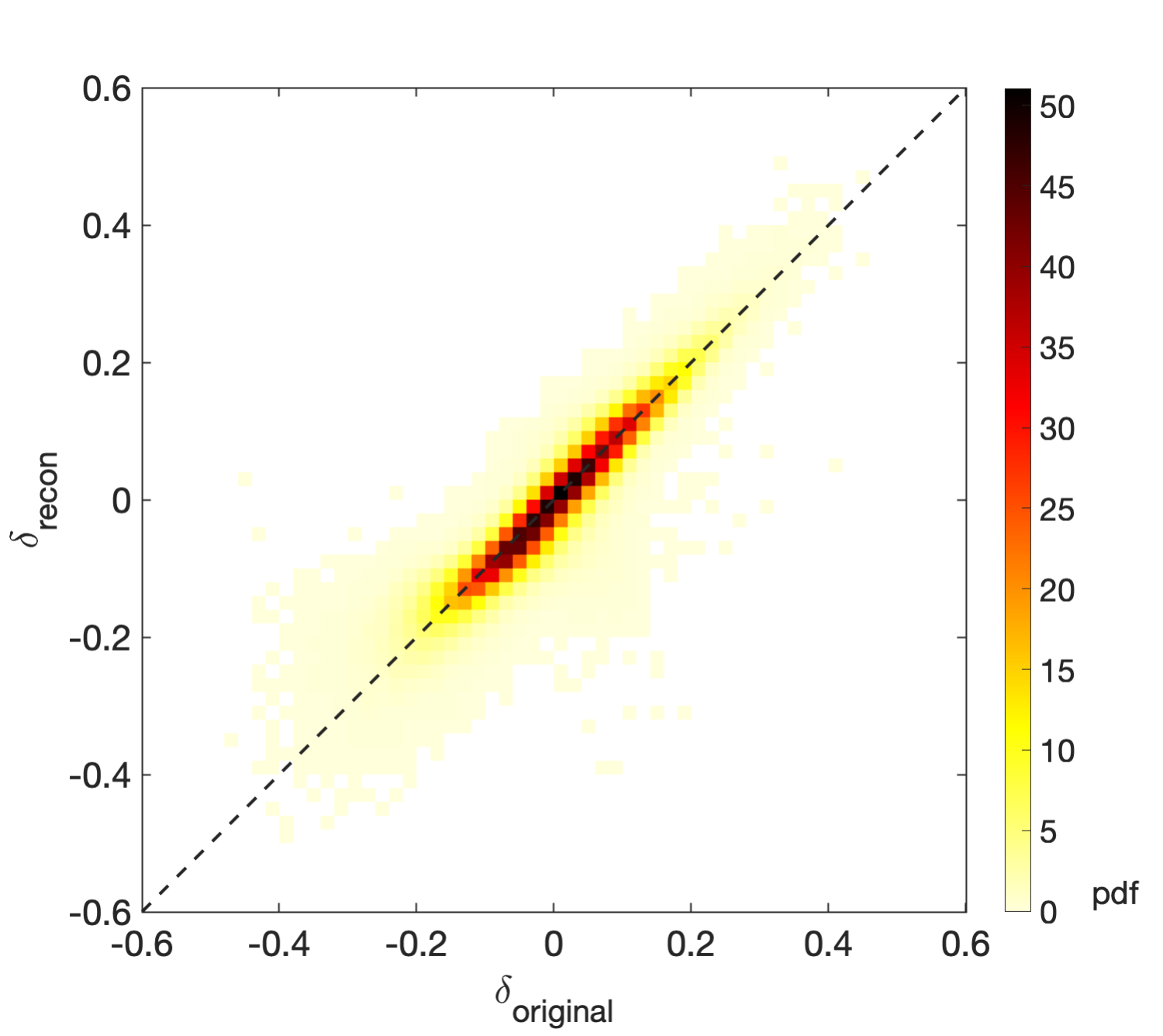}
    \includegraphics[width= 0.66\columnwidth]{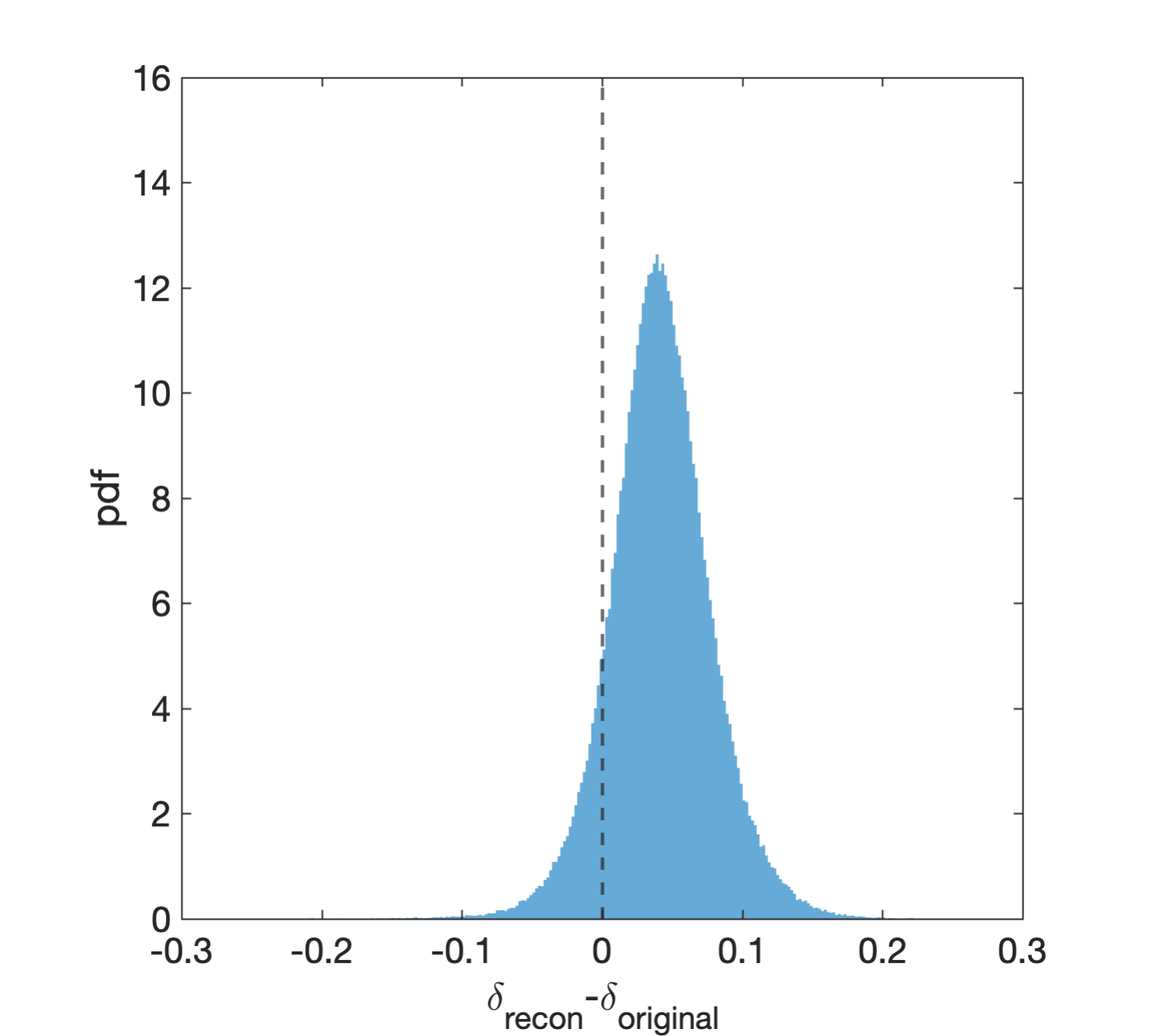}
    \includegraphics[width= 0.66\columnwidth]{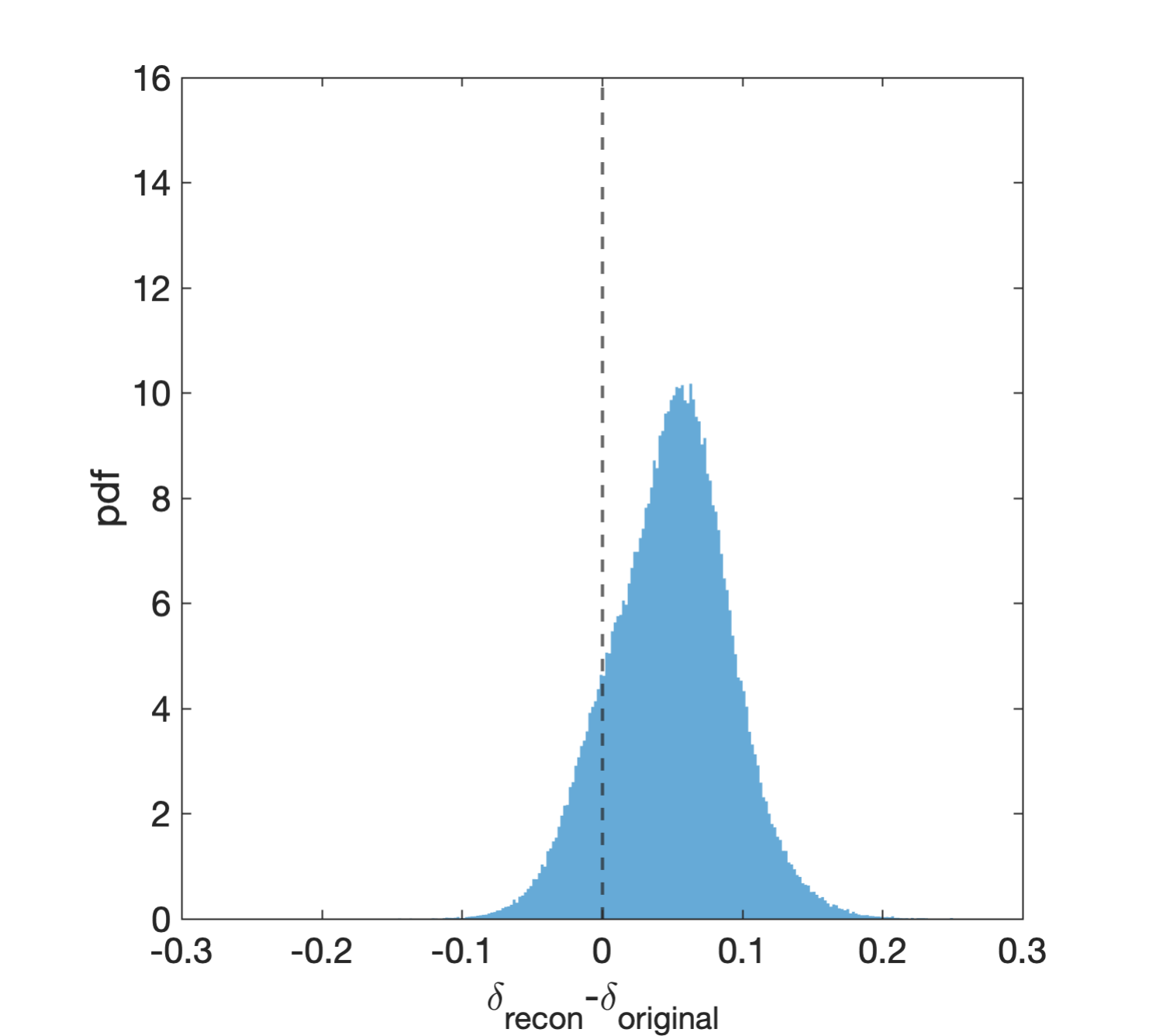}
    \includegraphics[width= 0.66\columnwidth]{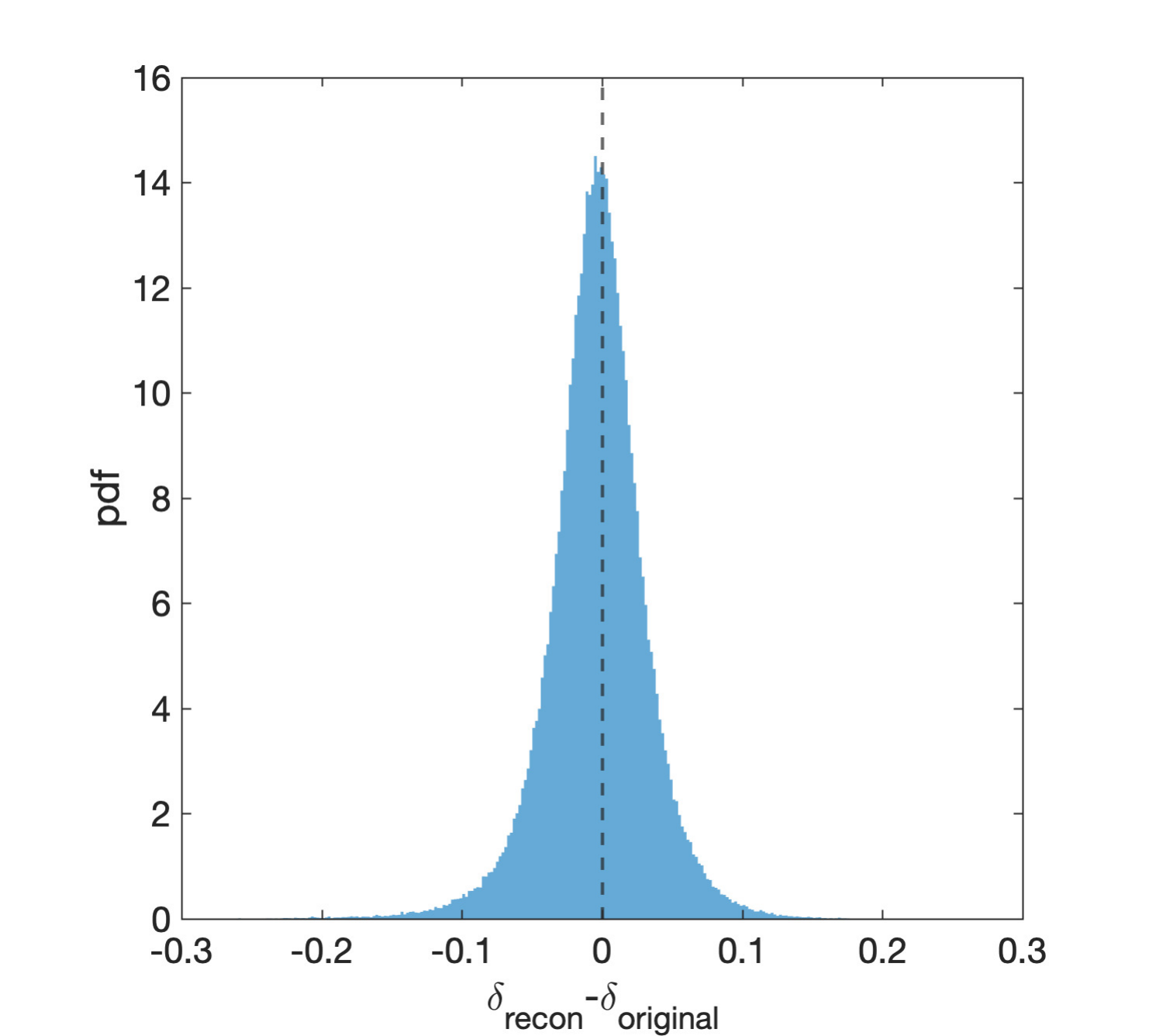}\\
\caption{Calibration of the initial density reconstruction. (Top) we show the reconstructed initial overdensity (``$\delta_{\rm recon}$'') vs the input, true one (``$\delta_{\rm original}$''). (Bottom) the PDF of the residual $\delta_{\rm recon}-\delta_{\rm original}$. The reconstruction is made using the mock observations of the 21~cm and CO maps in a coeval box (from left to right) at redshift $z=7.56$, $8.20$, and $9.54$, respectively. The dashed lines indicate the perfect matching.} 
    \label{fig:histogram_residual}
\end{figure*}

\begin{figure*}
    \centering
    \includegraphics[width=0.5\columnwidth]{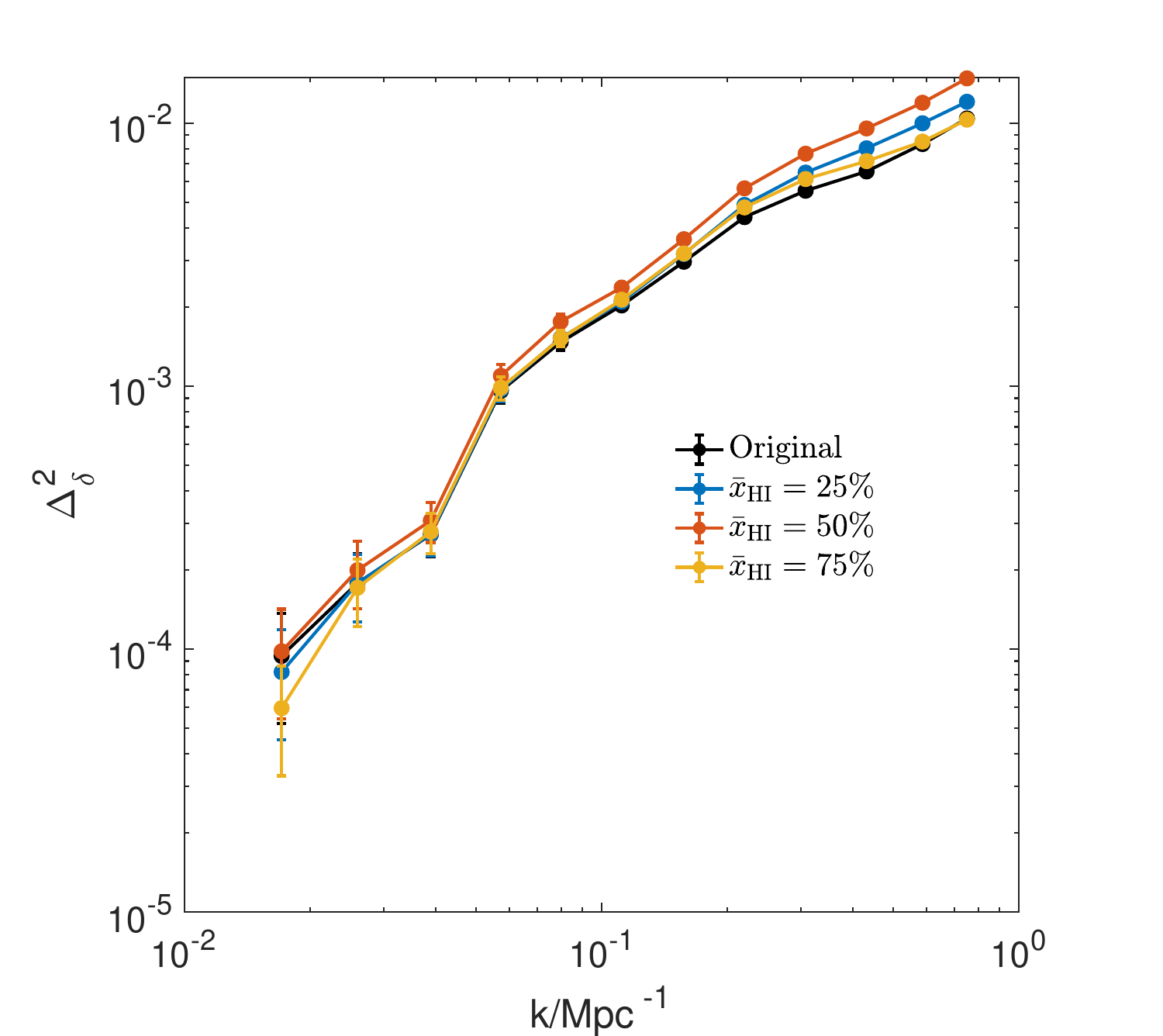}
    \includegraphics[width=0.5\columnwidth]{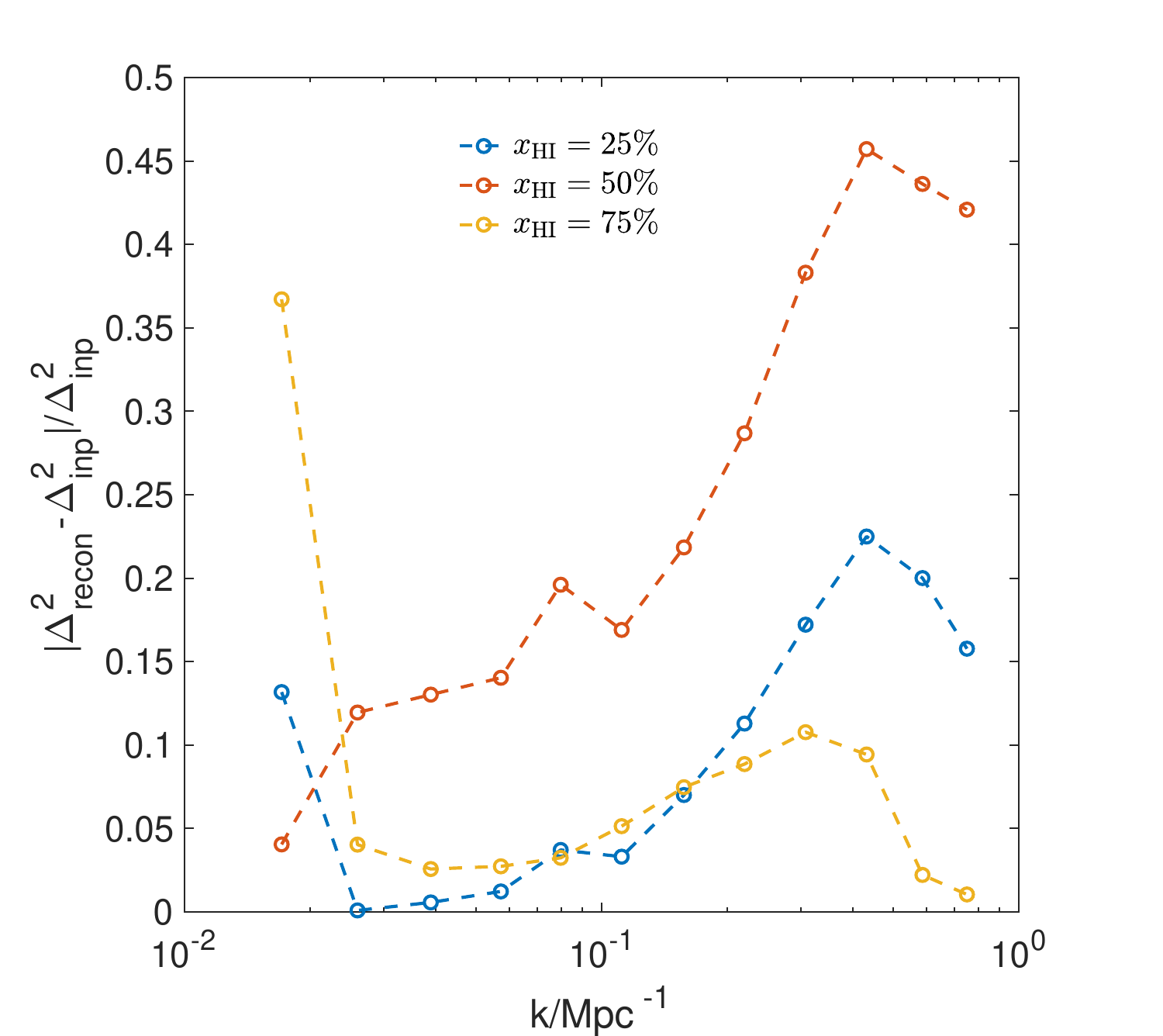}
    \includegraphics[width=0.5\columnwidth]{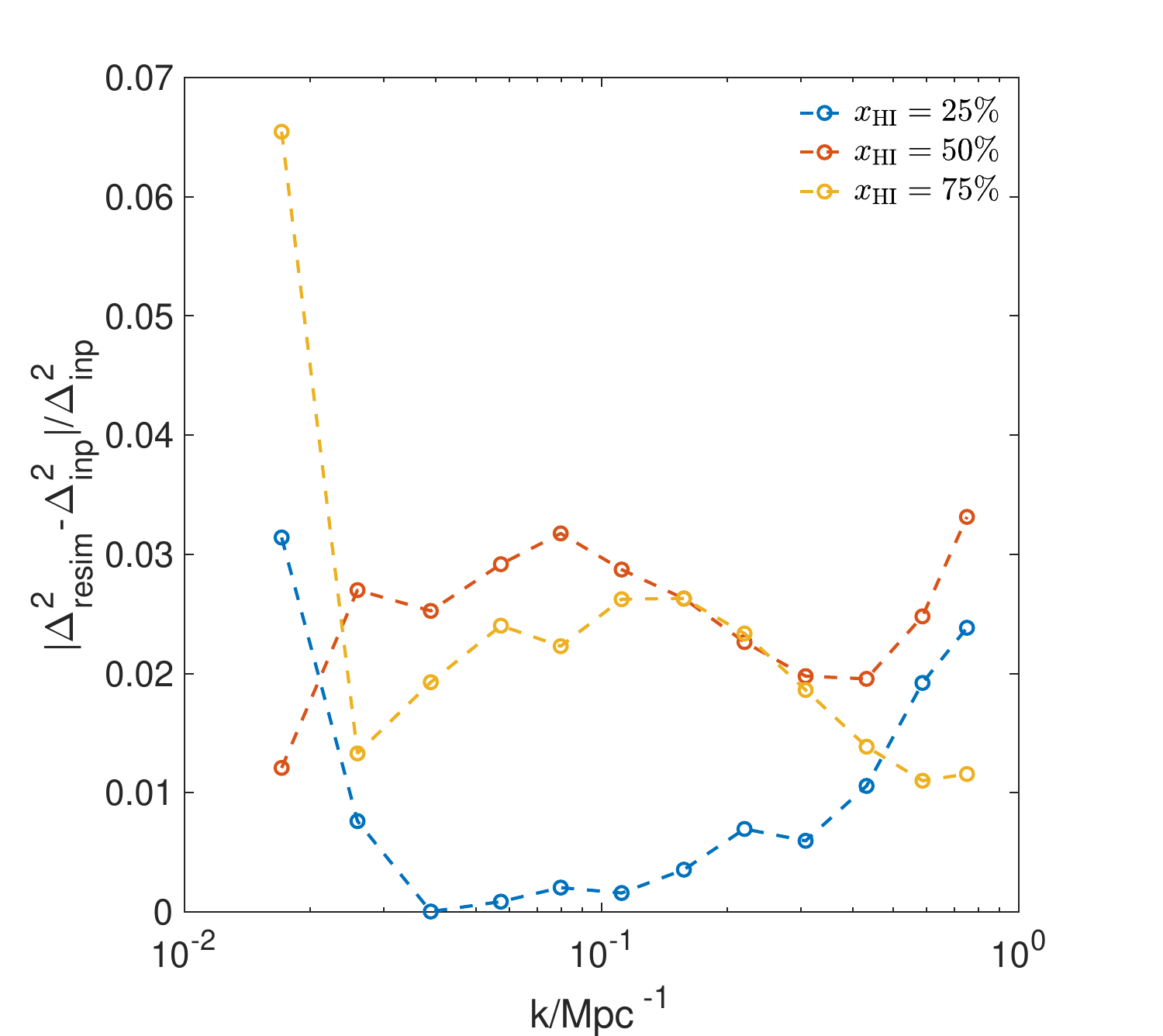}
    \includegraphics[width=0.5\columnwidth]{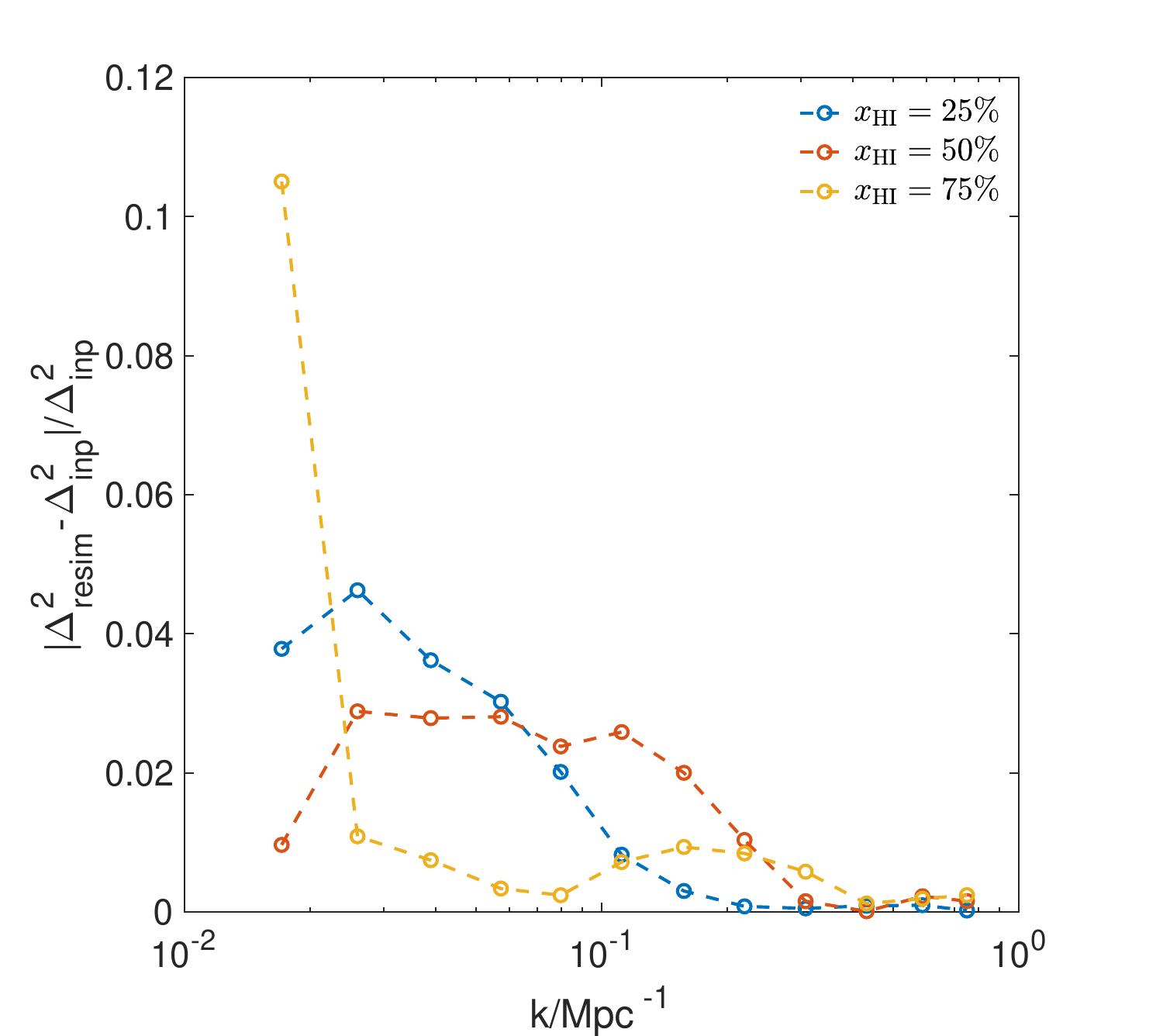}
    \caption{(First panel from the left) the dimensionless overdensity power spectrum $\Delta^2_\delta$ as a function of wavenumber $k$. We show the input, true overdensity power spectrum (black) and that reconstructed from the mock observations of the 21~cm and CO maps at different stages of reionization $\bar{x}_{\rm HI} = 0.25$ (blue), $0.50$ (red) and $0.75$ (yellow), respectively. (Second panel) the fractional difference of the reconstructed overdensity power spectrum with respect to the true power spectrum. 
    (Third panel) the fractional difference of the power spectrum of the resimulated 21~cm map with respect to that of the input 21~cm map. 
    (Fourth panel) same as the third panel but for the CO map.}
    \label{fig:ps_comparison}
\end{figure*}

\begin{figure}
    \centering
    \includegraphics[width=0.9\columnwidth]{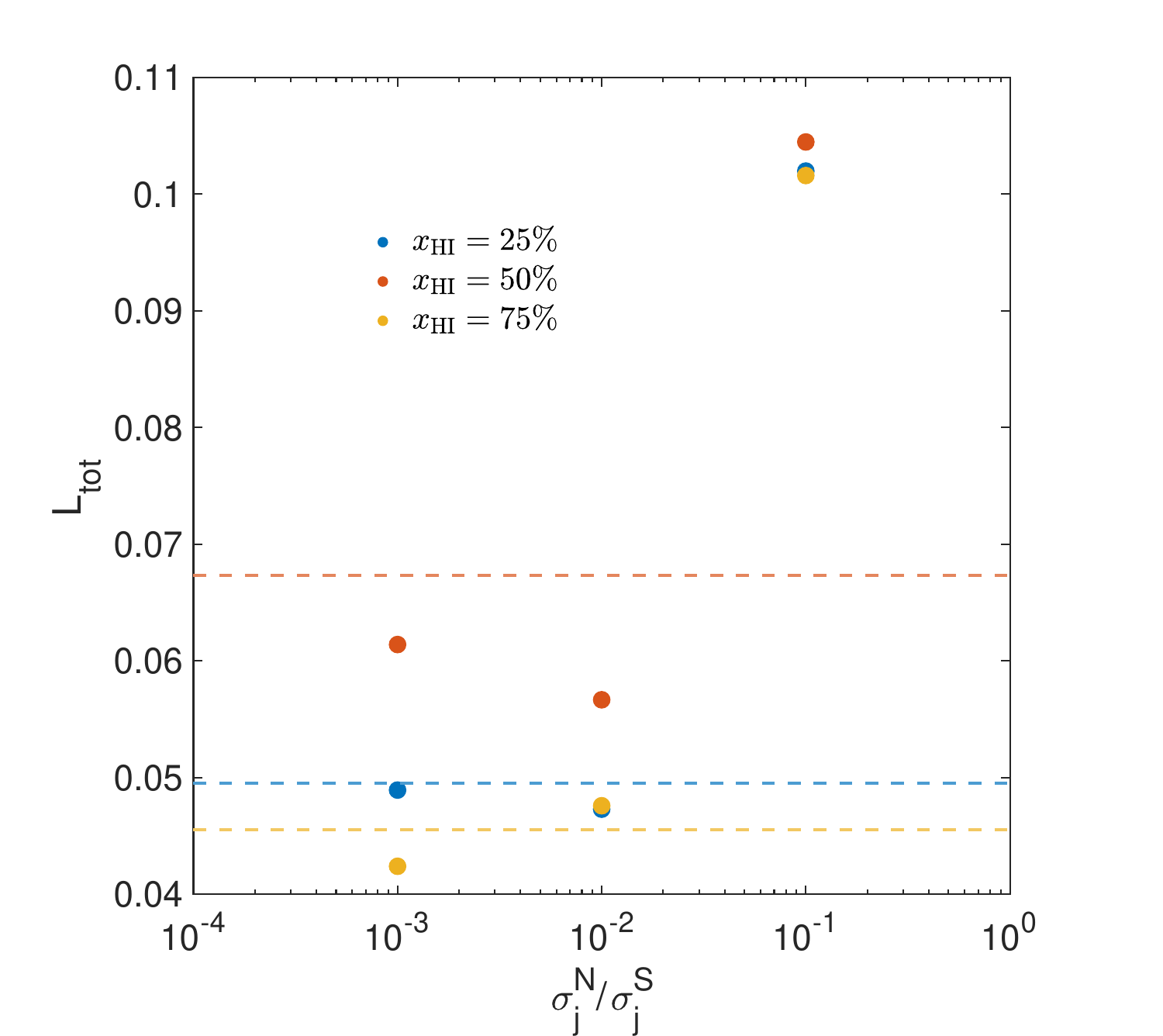}
    \caption{The goodness of reconstruction $L_{\rm tot}$ as a function of the ratio between the noise and the standard deviation of the signal map, $\sigma^{\rm N}_j/\sigma^{\rm S}_{j}$. We show the reconstruction from the mock observations of the 21~cm and CO maps at different stages of reionization $\bar{x}_{\rm HI} = 0.25$ (blue), $0.50$ (red) and $0.75$ (yellow), respectively. The dashed lines indicate the results for the ideal case without noise. Note that each case has only one realization, so the cases with small noise $\sigma^{\rm N}_j/\sigma^{\rm S}_{j}  \lesssim 10^{-2}$ may have even better results than the ideal case.   }  
    \label{fig:best-thermal-vs-L}
\end{figure}

\section{Results}
\label{result}

We present the optimal performance of the reconstruction in this section while leaving the optimization technique to Appendix~\ref{strategy}. 

\subsection{Mock with Cosmic Signals}

We first consider the reconstruction in the ideal case with only cosmic signals and no noise. From the reconstructed initial density, we resimulate the 21~cm and CO intensity maps at given redshifts. In Table~\ref{table of uncertainty}, we show the normalized rms difference, $L_{\rm tot}$, as a metric for evaluating the goodness of reconstruction, which is defined as 
 \begin{eqnarray}
     L_{\rm tot} = \left[\frac{1}{2N_p}\sum_{j,\,\alpha}\frac{\left(T^{\rm{mod}}_{j,\,\alpha}-T^{\rm inp}_{j,\,\alpha}\right)^2}{(\sigma^{\rm S}_{j})^{2}} \right]^{1/2}\,,
     \label{eqn:Ltot}
 \end{eqnarray}
 where $N_p$ is the number of coarse-grained cells in mock observations, and $\sigma^{\rm S}_{j}$ is the standard deviation of the ``signal maps'', i.e.\ the mock maps that include only cosmic signals and no noise. Note that we use $\sigma^{\rm S}_{j}$, instead of the amplitude of the input temperature itself, as the denominator in Equation~(\ref{eqn:Ltot}), because the temperature can be nearly zero in some points but the field amplitude can be represented by its standard deviation $\sigma^{\rm S}_{j}$ statistically. We find that the reconstruction is accurate in the sense that the errors in the resimulated 21~cm or CO maps are within $7\%$ on average with respect to the input maps. 

For visualization purposes, we show a slice of the resimulated 21~cm intensity maps and the input, mock observation in Figure~\ref{fig:21cm_images}, and the same comparison for the CO intensity maps in Figure~\ref{fig:CO_images}. Comparison by eye finds almost no difference between the resimulated map and the input map. In Figure~\ref{fig:image_comparison_pure}, we compare a slice of the input, true initial overdensity with the reconstructed initial overdensity field using the mock observations from three different redshifts. Their difference is small as seen in the bottom panel, too. 

To evaluate the difference in a quantitative manner, we plot the comparison of overdensity in the top panel of Figure~\ref{fig:histogram_residual}, and the probability distribution function (PDF) of the residual overdensity in the bottom panel of Figure~\ref{fig:histogram_residual}. We find that the distribution of the residual overdensity is Gaussian. At high redshift $z=9.54$ ($\bar{x}_{\rm HI} = 0.75$), the distribution has a zero mean, which implies that the reconstruction is unbiased. However, at lower redshifts $z=8.20$ and $7.56$ ($\bar{x}_{\rm HI} = 0.50$ and $0.25$ respectively), the mean of the distribution is on the positive side, which indicates that the reconstructed initial overdensity is overestimated with respect to the true initial overdensity. Also, the top panel shows that this overestimation mostly takes place in the overdense regions ($\delta >0$). This is likely due to the fact that at lower redshifts, the ionized bubbles have large sizes ($\sim$ tens of Mpc), so the impact of the initial density field on the 21~cm maps at the later stage of reionization is not as local as that in the early stage of reionization. 

To further evaluate the clustering of the reconstructed initial density field, we plot its power spectrum in the first panel of Figure~\ref{fig:ps_comparison}, and its fractional difference with respect to the input, true initial power spectrum in the second panel of Figure~\ref{fig:ps_comparison}. While we find their agreement at large scales, the reconstructed overdensity power spectrum is overestimated at small scales, particularly at the middle stage of reionization ($\bar{x}_{\rm HI} = 0.50$) at the level of error $\sim$ tens of percent. 

Regarding the power spectrum of the 21~cm and CO maps, the power spectra of the resimulated maps and the input maps are almost indistinguishable, so we only plot their fractional difference in the third and fourth panel of Figure~\ref{fig:ps_comparison}. We find that the power spectra of the resimulated map have an error of $\lesssim 4\%$ with respect to the true power spectra in most cases. 


\subsection{Mock with Noises}

We estimate the effect of noise on the reconstruction in this subsection. Again, as a proof of concept, here we assume a white noise that sums up all noises in observations and leave it to follow-up work to include more realistic modeling of noises and systematics. We include a noise $\sigma^{\rm N}_j$ in the input mock map, with three scenarios in terms of its ratio to the standard deviation of the ``signal map'' $\sigma^{\rm S}_{j}$: $\sigma^{\rm N}_j/\sigma^{\rm S}_{j} = 10^{-3}$, $10^{-2}$, and $10^{-1}$. In Figure~\ref{fig:best-thermal-vs-L}, we plot the goodness of reconstruction as a function of $\sigma^{\rm N}_j/\sigma^{\rm S}_{j}$. We find that the goodness of reconstruction is as good as the ideal case without noise if $\sigma^{\rm N}_j/\sigma^{\rm S}_{j} \lesssim 0.01$. However, once the noise is at the level of $0.1\,\sigma^{\rm S}_{j}$, the reconstruction is significantly worse (with rms error $\gtrsim 0.1$) than the ideal case. This can be considered as a rough estimation of the noise level required for future observations for the purpose of the initial density reconstruction.



\section{Conclusion}
\label{conclusion}

In this paper, we propose to reconstruct the cosmological initial density field using the \hi\ 21~cm and CO line intensity maps from the EoR. We employ the conjugate gradient method and develop the machinery for minimizing the cost function for the intensity mapping observations and apply this framework to the reconstruction from the EoR observations. Specifically, an analytical formalism for the gradient of the cost function is derived using the ESMR and the Zel'dovich approximation as the underlying theory for reionization and density fluctuations. Our results demonstrate that the resimulated intensity maps match the input maps of mock observations with an rms error $\lesssim 7\%$ at all stages of reionization. This reconstruction is also robust at the same level of accuracy when a noise at the level of $\lesssim 1\%$ of the standard deviation of the signal map is applied to each map. This suggests that our work provides an effective technique for reconstructing the cosmological initial density distribution from high-redshift observations.

Nevertheless, our proof-of-concept work has a few limitations. We only adopt a simple treatment of smoothing kernels from the simulation cells to the observation pixels, and assume a white noise that sums up all noises in observations. In principle, our work can be extended to include more realistic modeling of smoothing, noises, and observational effects (e.g.\ redshift-space distortions) as parts of forward simulations, which will be further developed in a follow-up paper. 


\section*{Acknowledgements}
This work is supported by the National SKA Program of China (grant No.~2020SKA0110401) and NSFC (grant No.~11821303). We thank Adrian Liu and Richard Grumitt for the useful discussions. We acknowledge the Tsinghua Astrophysics High-Performance Computing platform at Tsinghua University for providing computational and data storage resources that have contributed to the research results reported within this paper. 

\begin{appendix}

\begin{figure*}
    \centering
    \includegraphics[width=0.67\columnwidth]{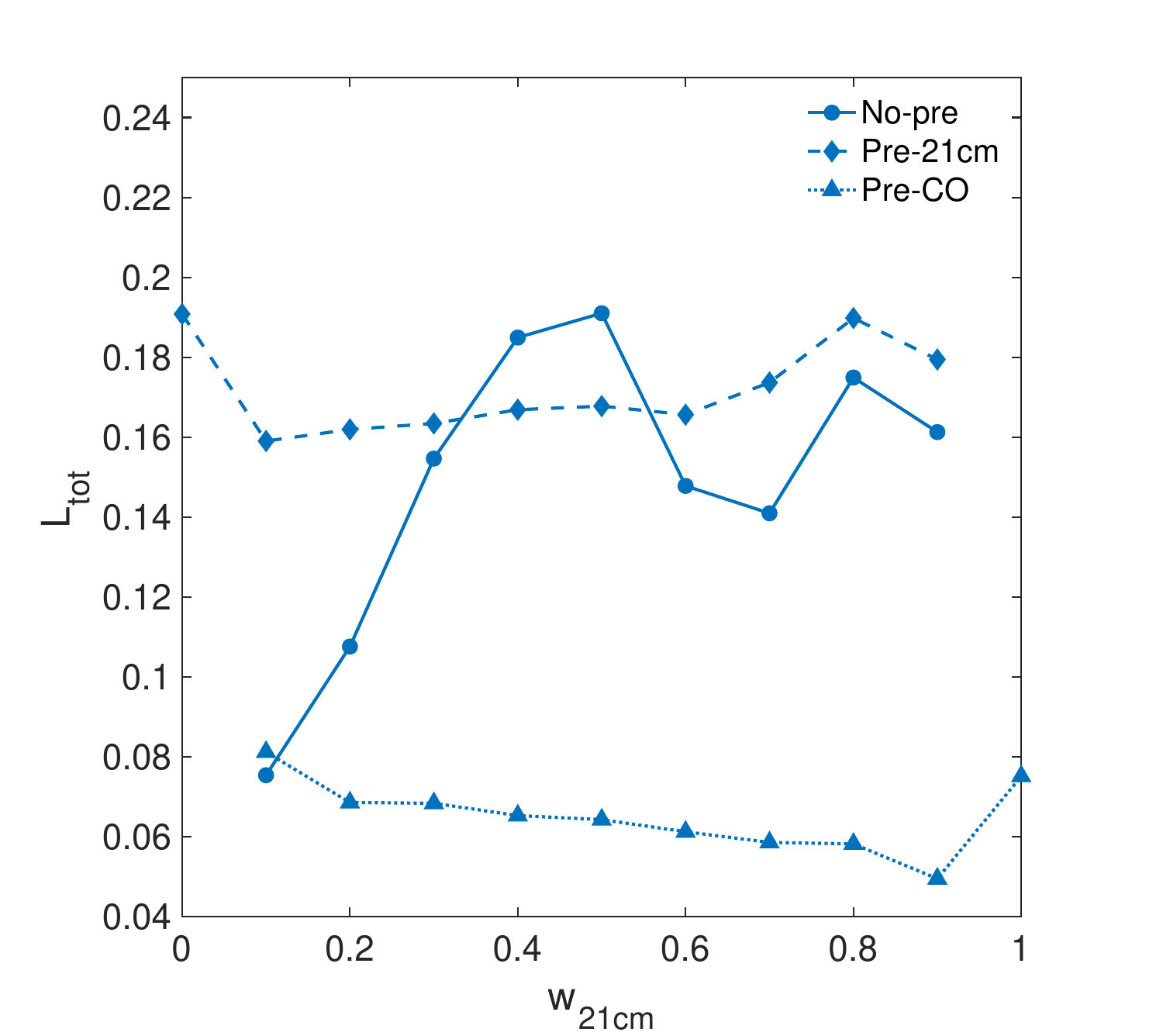}
    \includegraphics[width=0.67\columnwidth]{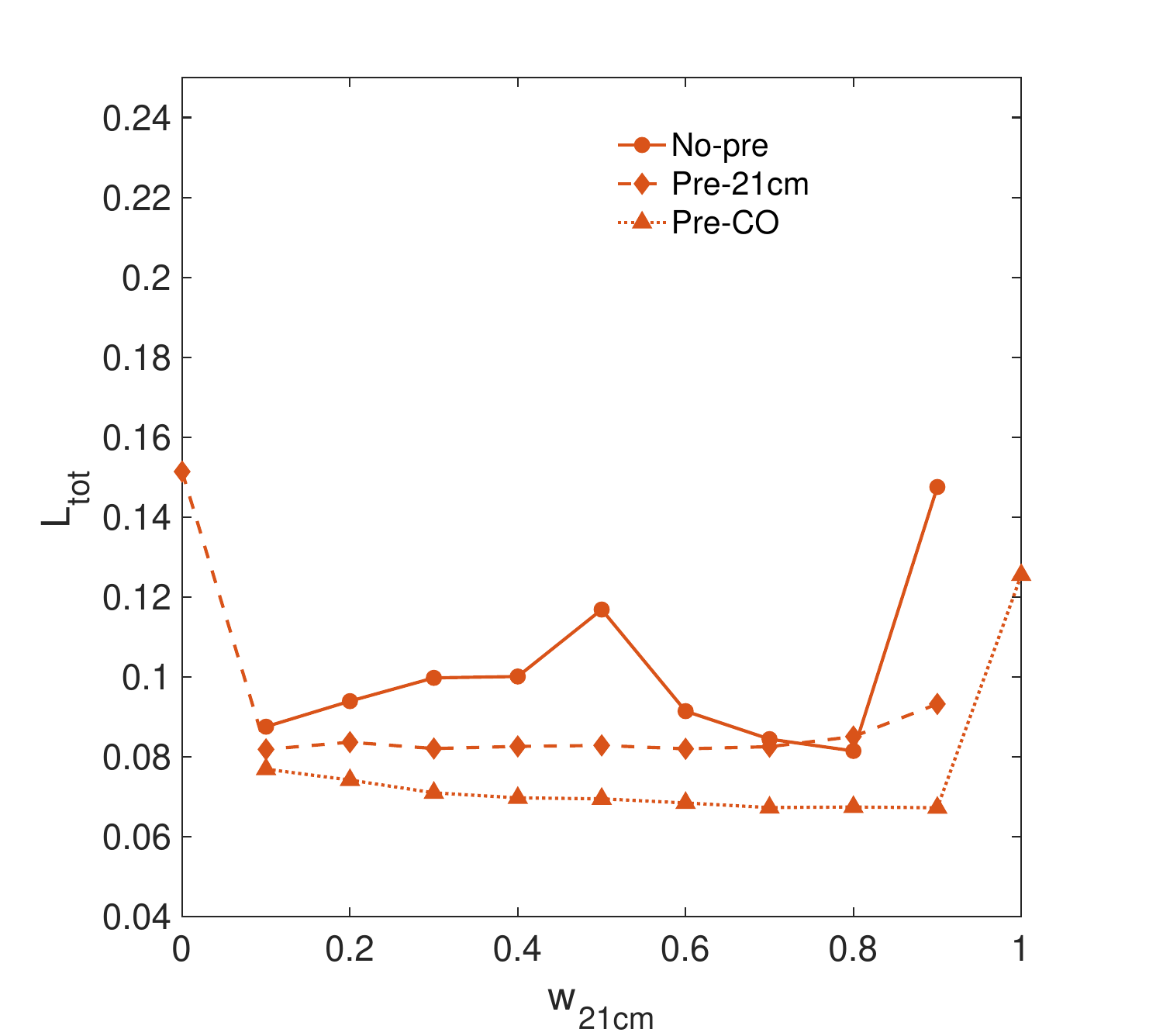}
    \includegraphics[width=0.67\columnwidth]{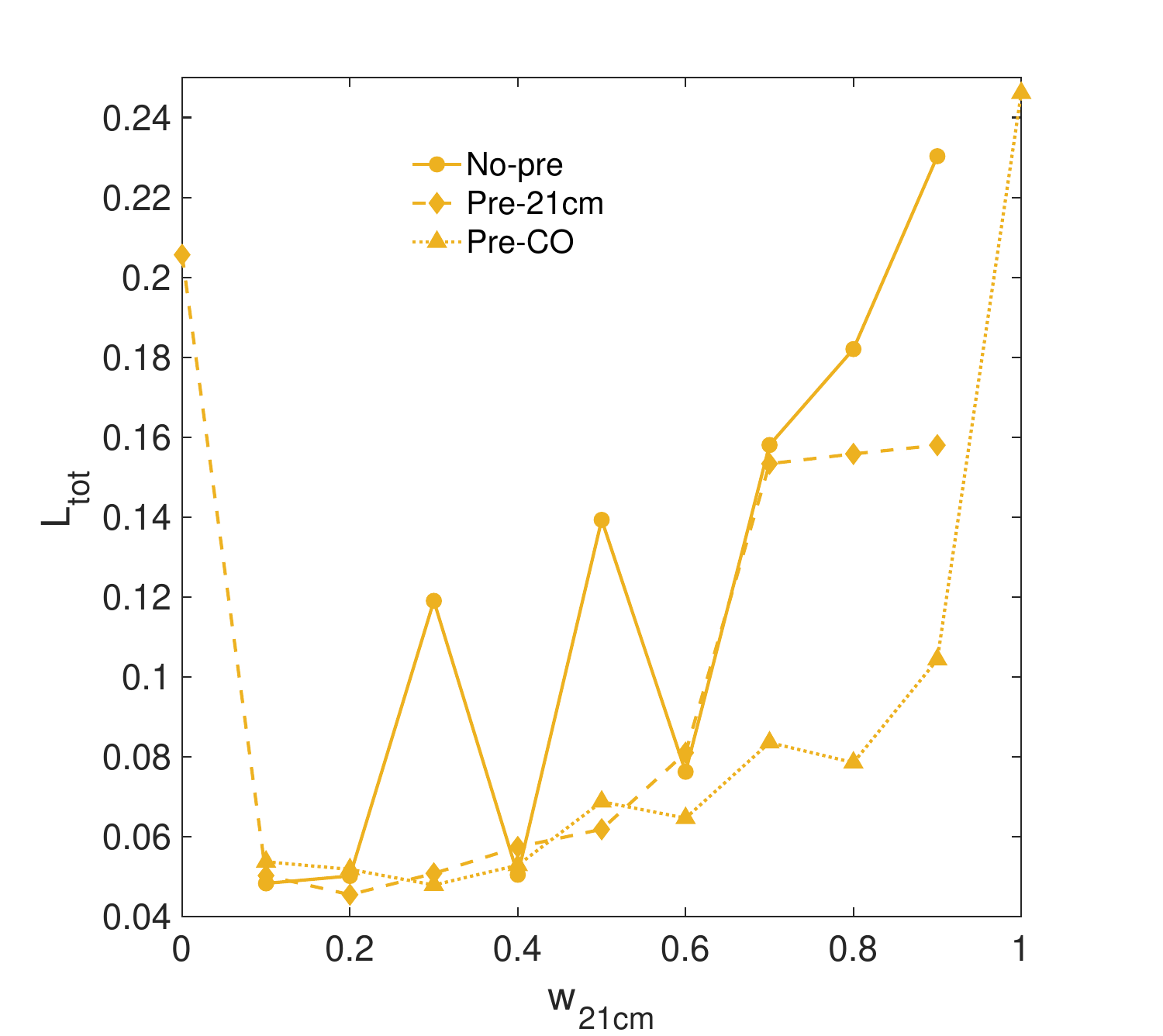}
    \caption{The goodness of reconstruction $L_{\rm tot}$ vs the weight $w_{\rm 21cm}$ in the step of full construction for different strategies --- ``No-pre'' (dots), ``Pre-21cm'' (diamonds) and ``Pre-CO'' (triangles) --- (from left to right) at redshift $z=7.56$, $8.20$ and $9.54$, corresponding to $\bar{x}_{\rm HI} = 0.25$, $0.50$ and $0.75$, respectively.} 
    \label{fig:weight-vs-L}
\end{figure*}

\begin{table*}
    \centering
    \caption{Optimal Performance of Reconstruction Strategies.}
     \begin{tabular}{ccccclcccclccccl}
     \hline \hline 
                 &  \multicolumn{5}{c}{$\bar{x}_{\rm HI} = 0.25$}  &  \multicolumn{5}{c}{$\bar{x}_{\rm HI} = 0.50$} &  \multicolumn{5}{c}{$\bar{x}_{\rm HI} = 0.75$} \\ 
			 \cmidrule(l{.75em}l{.75em}r{.75em}){2-6}
			 \cmidrule(l{.75em}l{.75em}r{.75em}){7-11}
			 \cmidrule(l{.75em}l{.75em}r{.75em}){12-16}
   Strategy & $w_{\rm 21cm}$ & $N_{\rm pre}$ &  $N_{\rm full}$ & $t_{\rm tot}$ [hrs] & $L_{\rm tot}$ & $w_{\rm 21cm}$ & $N_{\rm pre}$ &  $N_{\rm full}$ & $t_{\rm tot}$ [hrs] & $L_{\rm tot}$ & $w_{\rm 21cm}$ & $N_{\rm pre}$ &  $N_{\rm full}$ & $t_{\rm tot}$ [hrs] & $L_{\rm tot}$ \\
   \hline
   No-pre     & 0.1 & ---   & 280 & 26  & 0.0754 & 0.8 & ---   & 100 & 10  & 0.0878  & 0.1 & ---  & 570  & 44 & 0.0484 \\ \hline
   Pre-21cm & 0.1 & 40  & 260 & 25  & 0.159  & 0.1 & 60   & 470 & 45 & 0.0819  & 0.2 & 100 & 540 & 53 & 0.0455* \\ \hline
   Pre-CO    & 0.9 & 600 &150 & 29  & 0.0495* & 0.9 & 500  & 70  & 18 & 0.0673*  & 0.3 & 600 & 300 & 39 & 0.0480 \\ \hline
    \end{tabular}
     \flushleft
    \tablenotetext{}{Note. --- For each strategy at a given stage of reionization (labeled by $\bar{x}_{\rm HI}$), we show the optimum weight $w_{\rm 21cm}$ in the step of full construction, the number of iterations in the preprocessing $N_{\rm pre}$, the number of iterations in the full reconstruction $N_{\rm full}$, the total wall-clock time $t_{\rm tot}$ based on a test using eight CPU cores, and the goodness of reconstruction $L_{\rm tot}$. The star (`*') marks the optimal choice of strategy and weight for each mock observation at a given $\bar{x}_{\rm HI}$, which is used in the main text of this paper.}
    \label{strategy summary}
\end{table*}

\section{Optimization of Reconstruction}
\label{strategy}

In this section, we discuss the optimization technique of reconstruction. Numerically, the coefficients $w_{\rm 21cm}$ and $w_{\rm CO}$ in Equation~(\ref{cost}) can be adjusted to control the weights of the 21~cm and CO intensity mapping measurements on the cost function, respectively. Another trick is to introduce a preprocessing step --- reconstruction with only one type of observation --- followed by the full reconstruction (i.e.\ with both types of measurements simultaneously). The preprocessing step obtains an initial guess closer to the minimum than a random guess, with moderate, yet additional computational costs. As such, we consider three strategies --- ``no-pre'' (no preprocessing and directly full reconstruction), ``pre-21cm'' (preprocessing with the 21~cm maps, followed by the full reconstruction), and ``pre-CO'' (preprocessing with the CO maps, followed by the full reconstruction). For each strategy, we also vary the weight $w_{\rm 21cm}$ in the step of full reconstruction. Note that $w_{\rm CO} = 1-w_{\rm 21cm}$. For simplicity, we neglect the noise in this section. 

Figure~\ref{fig:weight-vs-L} shows that for the mock observations at $\bar{x}_{\rm HI}=0.25$ and $0.50$, the ``pre-CO'' strategy performs the best in terms of the smallest $L_{\rm tot}$. For the mock observations at $\bar{x}_{\rm HI}=0.75$, the optimums for these three strategies are very close, with the ``pre-21cm'' strategy slightly better. In Table~\ref{strategy summary}, we list the performance results at the respective optimum for each strategy at a given stage of reionization, in a test with eight cores of Intel Xeon E78860V3 2.20GHz CPU. While the wall-clock time of the optimum case (as marked by the star `*') is not the smallest for each mock observation, Table~\ref{strategy summary} shows that the optimum is a good trade-off between precision and efficiency.

\section{Subtraction of mean signals}
\label{app:mean}

For the interferometric array observations, the ${\bf k}_\perp = 0 $ (zero baseline) mode is not observable so the mean is subtracted from the signal. In this section, we discuss whether this changes our reconstruction. 

In a coeval box, the subtracted signal at each frequency channel is $    \Delta T_\alpha = T_\alpha - \frac{1}{N_c}\sum_\beta T_\beta$, where $N_c$ is the total number of pixels on the two-dimensional map at each frequency channel. If we use the subtracted signal to evaluate the cost function and gradients, Equation~(\ref{likelihood derivative3}) is changed as 
\begin{eqnarray}
    \frac{\partial \phi}{\partial T_{j,\,\alpha}^{\rm mod}}  &=& C_{\rm cost} \,w_{j}\, \biggl\lbrack \frac{ (T^{\rm{mod}}_{j,\,\alpha}-T^{\rm inp}_{j,\,\alpha})}{(\sigma^{\rm N}_{j})^2}\nonumber\\
    & & \!\!\!\!\!\!\!\!\!\!\!\!\!\!\!\!\!\!\!\!\!\!\!\!\!\!\!\!\!\!\!\!\!\!\!\!   - \frac{1}{N_c}\sum_\beta\frac{(T^{\rm{mod}}_{j,\,\beta}-T^{\rm inp}_{j,\,\beta})}{(\sigma^{\rm N}_{j})^2} -\frac{1}{N_c}\sum_\beta\frac{(\Delta T^{\rm{mod}}_{j,\,\beta}-\Delta T^{\rm inp}_{j,\,\beta})}{(\sigma^{\rm N}_{j})^2} \biggr\rbrack \,.
\label{eqn:subtraction}
\end{eqnarray}

The third term on the RHS of Equation~(\ref{eqn:subtraction}) is zero statistically in coeval boxes and in a lightcone as it is the mean of subtracted signals at each frequency channel. The second term on the RHS of Equation~(\ref{eqn:subtraction}) is the difference of mean signals. It may be different from the full signal case at the first few steps if fluctuations of our initial guess are smaller than the true field. Nevertheless, this effect becomes small once beyond the burn-in phase. Therefore, we conclude that for the reconstruction of initial density, the effect of subtracting the mean is negligible.

\end{appendix}
\bibliographystyle{aasjournal}
\bibliography{recon}

\begin{thebibliography}{}
\expandafter\ifx\csname natexlab\endcsname\relax\def\natexlab#1{#1}\fi
\providecommand{\url}[1]{\href{#1}{#1}}
\providecommand{\dodoi}[1]{doi:~\href{http://doi.org/#1}{\nolinkurl{#1}}}
\providecommand{\doeprint}[1]{\href{http://ascl.net/#1}{\nolinkurl{http://ascl.net/#1}}}
\providecommand{\doarXiv}[1]{\href{https://arxiv.org/abs/#1}{\nolinkurl{https://arxiv.org/abs/#1}}}

\bibitem[{{Bernal} \& {Kovetz}(2022)}]{2022A&ARv..30....5B}
{Bernal}, J.~L., \& {Kovetz}, E.~D. 2022, \aapr, 30, 5,
  \dodoi{10.1007/s00159-022-00143-0}

\bibitem[{Brent(1973)}]{Brent1973}
Brent, R.~P. 1973, {Algorithms for Minimization without Derivatives}, 1st edn.
  (Englewood Cliffs, New Jersey: Prentice-Hall)

\bibitem[{{Chang} {et~al.}(2015){Chang}, {Gong}, {Santos}, {Silva}, {Aguirre},
  {Dor{\'e}}, \& {Pritchard}}]{2015aska.confE...4C}
{Chang}, T.~C., {Gong}, Y., {Santos}, M., {et~al.} 2015, in Advancing
  Astrophysics with the Square Kilometre Array (AASKA14), 4.
\newblock \doarXiv{1501.04654}

\bibitem[{{Chen} {et~al.}(2023){Chen}, {Mo}, \& {Wang}}]{2023MNRAS.526.2542C}
{Chen}, Y., {Mo}, H.~J., \& {Wang}, K. 2023, \mnras, 526, 2542,
  \dodoi{10.1093/mnras/stad2866}

\bibitem[{{Datta} {et~al.}(2014){Datta}, {Jensen}, {Majumdar}, {Mellema},
  {Iliev}, {Mao}, {Shapiro}, \& {Ahn}}]{2014MNRAS.442.1491D}
{Datta}, K.~K., {Jensen}, H., {Majumdar}, S., {et~al.} 2014, \mnras, 442, 1491,
  \dodoi{10.1093/mnras/stu927}

\bibitem[{{Datta} {et~al.}(2012){Datta}, {Mellema}, {Mao}, {Iliev}, {Shapiro},
  \& {Ahn}}]{2012MNRAS.424.1877D}
{Datta}, K.~K., {Mellema}, G., {Mao}, Y., {et~al.} 2012, \mnras, 424, 1877,
  \dodoi{10.1111/j.1365-2966.2012.21293.x}

\bibitem[{{Dumitru} {et~al.}(2019){Dumitru}, {Kulkarni}, {Lagache}, \&
  {Haehnelt}}]{2019MNRAS.485.3486D}
{Dumitru}, S., {Kulkarni}, G., {Lagache}, G., \& {Haehnelt}, M.~G. 2019,
  \mnras, 485, 3486, \dodoi{10.1093/mnras/stz617}

\bibitem[{{Furlanetto} {et~al.}(2004){Furlanetto}, {Zaldarriaga}, \&
  {Hernquist}}]{2004ApJ...613....1F}
{Furlanetto}, S.~R., {Zaldarriaga}, M., \& {Hernquist}, L. 2004, \apj, 613, 1,
  \dodoi{10.1086/423025}

\bibitem[{{Gong} {et~al.}(2012){Gong}, {Cooray}, {Silva}, {Santos}, {Bock},
  {Bradford}, \& {Zemcov}}]{2012ApJ...745...49G}
{Gong}, Y., {Cooray}, A., {Silva}, M., {et~al.} 2012, \apj, 745, 49,
  \dodoi{10.1088/0004-637X/745/1/49}

\bibitem[{{Gong} {et~al.}(2011){Gong}, {Cooray}, {Silva}, {Santos}, \&
  {Lubin}}]{2011ApJ...728L..46G}
{Gong}, Y., {Cooray}, A., {Silva}, M.~B., {Santos}, M.~G., \& {Lubin}, P. 2011,
  \apjl, 728, L46, \dodoi{10.1088/2041-8205/728/2/L46}

\bibitem[{{Jasche} \& {Wandelt}(2013)}]{2013MNRAS.432..894J}
{Jasche}, J., \& {Wandelt}, B.~D. 2013, \mnras, 432, 894,
  \dodoi{10.1093/mnras/stt449}

\bibitem[{{Lidz} {et~al.}(2011){Lidz}, {Furlanetto}, {Oh}, {Aguirre}, {Chang},
  {Dor{\'e}}, \& {Pritchard}}]{2011ApJ...741...70L}
{Lidz}, A., {Furlanetto}, S.~R., {Oh}, S.~P., {et~al.} 2011, \apj, 741, 70,
  \dodoi{10.1088/0004-637X/741/2/70}

\bibitem[{Mesinger {et~al.}(2011)Mesinger, Furlanetto, \&
  Cen}]{Mesinger201121cmfast}
Mesinger, A., Furlanetto, S., \& Cen, R. 2011, Mon. Not. R. Astron. Soc., 411,
  955

\bibitem[{{Mo} {et~al.}(2010){Mo}, {van den Bosch}, \&
  {White}}]{2010gfe..book.....M}
{Mo}, H., {van den Bosch}, F.~C., \& {White}, S. 2010, {Galaxy Formation and
  Evolution} (Cambridge: Cambride Univ. Press)

\bibitem[{Polak(1971)}]{polak1971computational}
Polak, E. 1971, Computational methods in optimization: a unified approach,
  Vol.~77 (Academic press)

\bibitem[{{Press} \& {Schechter}(1974)}]{1974ApJ...187..425P}
{Press}, W.~H., \& {Schechter}, P. 1974, \apj, 187, 425, \dodoi{10.1086/152650}

\bibitem[{Press {et~al.}(2007)Press, Teukolsky, Vetterling, \&
  Flannery}]{10.5555/1403886}
Press, W.~H., Teukolsky, S.~A., Vetterling, W.~T., \& Flannery, B.~P. 2007,
  Numerical Recipes 3rd Edition: The Art of Scientific Computing, 3rd edn.
  (USA: Cambridge University Press)

\bibitem[{{Sheth} \& {Tormen}(1999)}]{1999MNRAS.308..119S}
{Sheth}, R.~K., \& {Tormen}, G. 1999, \mnras, 308, 119,
  \dodoi{10.1046/j.1365-8711.1999.02692.x}

\bibitem[{{Silva} {et~al.}(2015){Silva}, {Santos}, {Cooray}, \&
  {Gong}}]{2015ApJ...806..209S}
{Silva}, M., {Santos}, M.~G., {Cooray}, A., \& {Gong}, Y. 2015, \apj, 806, 209,
  \dodoi{10.1088/0004-637X/806/2/209}

\bibitem[{{Wang} {et~al.}(2013){Wang}, {Mo}, {Yang}, \& {van den
  Bosch}}]{2013ApJ...772...63W}
{Wang}, H., {Mo}, H.~J., {Yang}, X., \& {van den Bosch}, F.~C. 2013, \apj, 772,
  63, \dodoi{10.1088/0004-637X/772/1/63}

\end{thebibliography}

\end{document}